\documentclass[superscriptaddress,onecolumn]{revtex4}
\usepackage{graphicx,epsfig}
\usepackage{amsmath}
\usepackage {amssymb}
\usepackage{multirow}
\usepackage{xfrac}
\usepackage{ulem}
\usepackage{float}
\usepackage{color}

\newcommand{\be}{\begin{eqnarray}}
\newcommand{\ee}{\end{eqnarray}}
\newcommand{\bea}{\begin{eqnarray}}

\newcommand{\eea}{\end{eqnarray}}

\makeindex

\begin{document}

\title{Massive scalar field perturbations of 4D de Sitter Einstein-Gauss-Bonnet black holes}



\author{Ram\'{o}n B\'{e}car}
\email{rbecar@uct.cl}
\affiliation{Departamento de Ciencias Matem\'{a}ticas y F\'{\i}sicas, Universidad Catolica de Temuco}

\author{P. A. Gonz\'{a}lez}
\email{pablo.gonzalez@udp.cl} \affiliation{Facultad de
Ingenier\'{i}a y Ciencias, Universidad Diego Portales, Avenida Ej\'{e}rcito
Libertador 441, Casilla 298-V, Santiago, Chile.}

\author{Eleftherios Papantonopoulos}
\email{lpapa@central.ntua.gr}
\affiliation{Physics Division, School of Applied Mathematical and Physical Sciences, National Technical University of Athens, 15780 Zografou Campus,
    Athens, Greece.}

\author{Yerko V\'asquez}
\email{yvasquez@userena.cl}
\affiliation{Departamento de F\'isica, Facultad de Ciencias, Universidad de La Serena,\\
Avenida Cisternas 1200, La Serena, Chile.}

\date{\today }

\begin{abstract}

We investigate the propagation of massive scalar fields in the background of four-dimensional Einstein-Gauss-Bonnet black holes with de Sitter (dS) asymptotics. Our study focuses on the various branches of quasinormal modes present in this background, employing the pseudospectral Chebyshev method and the third-order Wentzel-Kramers-Brillouin approximation. We identify that the introduction of the Gauss-Bonnet coupling constant $\alpha$ gives rise to three branches of modes: the perturbative (in $\alpha$) Schwarzschild branch, the perturbative (in $\alpha$) dS branch, 
and a non-perturbative  (in $\alpha$) dS branch. 
Our results show that the propagation of a massive scalar field is stable in this background. Furthermore, the Gauss-Bonnet coupling constant induces significant deviations in the Schwarzschild branch and smaller deviations in the perturbative dS branch compared to the corresponding branches in the Schwarzschild-dS limit. Additionally, the non-perturbative dS branch of modes, absent when $\alpha=0$, emerges as a novel feature of the Einstein-Gauss-Bonnet framework.

\end{abstract}

\maketitle
\flushbottom

\tableofcontents

\newpage

\section{Introduction}

The investigation of various modified theories of gravity have emerged in an effort to overcome certain intrinsic inconsistencies of General Relativity (GR)
and to address fundamental questions such as quantum gravity and the singularity problem. Recent observational results on dark matter and dark energy
require modifications of GR at short and large distances to generate a viable theory of gravity \cite{Joyce:2014kja,Barack_2018}. The recent detection of gravitational waves in the modern observational era
provides the opportunity to test gravitational theories in the strong-field regime and differentiate between them \cite{Abbott:2016blz,Abbott:2016nmj,Abbott:2017vtc,Abbott:2017oio,TheLIGOScientific:2017qsa}. The compact objects predicted by different modified theories
and the expected observational signatures can give us important intuition about their properties.

One of the most well-studied nontrivial extensions of GR is the Lovelock theory \cite{Lovelock}, which, apart from the Einstein-Hilbert term, also includes higher-order curvature terms. In fact, the Lovelock theory contains terms only up to the second-order derivatives in the equations of motion, which makes the theory more manageable. The simplest case of a second-order Lovelock theory is the Gauss-Bonnet theory (GB). The central issue of these theories is the stability of their black hole solutions. 
The stability was studied under
scalar, vector, and tensor perturbations \cite{Dotti:2004sh, Dotti:2005sq, Gleiser:2005ra, Konoplya:2008ix}. It was found that there exists a scalar mode instability in five dimensions, a tensor mode instability in six dimensions, and no instability in other dimensions. However, when considering dimensions higher than six, a fully consistent higher-dimensional theory requires the inclusion of higher-order Lovelock corrections. Incorporating these corrections, instabilities occur \cite{Konoplya:2017lhs}. In string theory higher-order curvature terms are known to appear introducing higher derivatives in the metric \cite{ Candelas}  leading to the appearance of perturbative ghosts. The second-order Lovelock theory taking the form of the GB combination \cite{Gross} has no higher derivatives in the effective string action, so we expect no ghosts to appear
in second-order.

It has long been recognized that GR predicts static and spherically symmetric black hole solutions in diverse spacetimes.  In four-dimensional spacetime, the GB term behaves as a total derivative, rendering it non-contributory to gravitational dynamics and thus precluding the existence of distinct black hole solutions. Given that our observed universe is four-dimensional, verifying the nature of GB gravity through astronomical observations presents an exceptionally challenging endeavor. To circumvent the stringent requirements of Lovelock’s theory and directly introduce the GB term into $4D$ gravity, four-dimensional Einstein-Gauss-Bonnet ($4D$-EGB) gravity has recently been reformulated taking the limit $D \rightarrow 4$ of its higher-dimensional version, where the coupling constant scales as $\alpha \rightarrow \frac{\alpha}{D-4}$ \cite{Glavan:2019inb}. The GB term shows a non-trivial contribution to the gravitational dynamics. The theory preserves the number of degrees of freedom and remains free from the Ostrogradsky instability.

The new theory has stimulated a series of recent research works concerning black hole solutions and the properties of the novel $4D$-EGB theory; for instance, spherically symmetric black hole solutions were discovered \cite{Glavan:2019inb}, generalizing Schwarzschild black holes and are also free from singularity. Additionally, charged black holes in the AdS spacetime \cite{Fernandes:2020rpa}, radiating black hole solutions \cite{Ghosh:2020syx}, an exact charged black hole surrounded by clouds of strings was investigated in \cite{Singh:2020nwo} and a static spherically symmetric noncommutative (NC) geometry inspired black hole solution with Gaussian mass distribution as a source was obtained in \cite{Ghosh:2020cob}. The generalization of these static black holes to the rotating case was also addressed \cite{Wei:2020ght},  in \cite{Gammon:2022bfu} the authors have constructed slowly rotating black hole solutions for $4D$-EGB gravity in asymptotically flat, de Sitter, and anti-de Sitter spacetimes.

On the other hand,  black holes and the generalization of the BTZ solution in the presence of higher curvature (GB and Lovelock) corrections of any order was found in Refs. \cite{Kumar:2020uyz,Yang:2020jno,Biswas:2022qyl,Jusufi:2022ukt,Kumar:2023ijg} and \cite{Hennigar:2020fkv,Konoplya:2020ibi}, respectively.  Furthermore, a $4D$ Einstein-Lovelock theory was formulated and black hole solutions were studied in \cite{Konoplya:2020qqh, Konoplya:2020der}. The intriguing physical properties of black holes in this $4D$-EGB gravity, including their thermodynamics, have been investigated \cite{ Hegde:2020cdm,HosseiniMansoori:2020yfj,Wei:2020poh,Singh:2020xju,EslamPanah:2020hoj,Ghaffarnejad:2021zbx,Belhaj:2022gcj,Belhaj:2022qmn,Ladghami:2023ccf}, Hawking radiation and greybody factors  \cite{Devi:2020uac,Zhang:2020qam, Konoplya:2020cbv}, quasinormal modes and stability \cite{Konoplya:2020bxa,Konoplya:2020juj, Mishra:2020gce,Ma:2022svl,Ladino:2023zqc,Vieira:2021doo, Zhang:2020sjh, Churilova:2020mif, Liu:2020evp, Aragon:2020qdc}, geodesic motion and shadow \cite{Guo:2020zmf,Heydari-Fard:2020sib,Roy:2020dyy,Chen:2021gwy,Liu:2022plm}, electromagnetic radiation from a thin accretion disk from spherically symmetric black holes \cite{Liu:2020vkh}, among others. In $4D$-EGB gravity, an instability arises when the coupling parameter, $\alpha$, is not sufficiently small. This instability manifests itself as a dynamical eikonal instability in both vector and scalar gravitational perturbations \cite{Konoplya:2020bxa,Konoplya:2020juj}, similar to those observed in the higher-dimensional EGB and Lovelock theories \cite{Dotti:2004sh, Dotti:2005sq, Gleiser:2005ra, Konoplya:2008ix,Konoplya:2017lhs, Konoplya:2017zwo}. Consequently, only small values of $\alpha$ are physically meaningful.

Despite the criticisms \cite{Gurses:2020ofy, Mahapatra:2020rds, Shu:2020cjw, Tian:2020nzb} about the approach applied in Ref. \cite{Glavan:2019inb}, that arises from the idea of defining a theory from a set of solutions that are obtained by the limit $D \rightarrow 4$ of the $D$-dimensional EGB theory,  have been proposed other approaches to obtain a well defined $D \rightarrow 4$ limit of the EGB theory in Refs. \cite{Aoki:2020lig, Aoki:2020iwm, Lu:2020iav, Fernandes:2020nbq, Hennigar:2020lsl}. In addition, an action with a set of field equations was found, using dimensional reduction methods \cite{Mann:1992ar,Lu:2020iav, Kobayashi:2020wqy}. The resulting theory corresponds to a scalar-tensor theory of Horndeski type. It is worth mentioning that all the solutions found in the original paper on the $4D$ -EGB theory \cite{Glavan:2019inb} were shown to be also solutions of the new formulation of the theory. In particular, the spherically symmetric Schwarzschild-like solution generated by this theory coincides with the metric of the $D \rightarrow 4$ limit of the $D$-dimensional EGB theory. So, it is important to emphasize that the spherically symmetric Schwarzschild-like solution obtained in the $D \rightarrow 4$ limit of the $D$-dimensional EGB theory is also a solution of theories formulated with a well defined limit $D \rightarrow 4$ of the EGB theory. 
Furthermore, it is worth noting that these black holes are also solutions of the semi-classical Einstein equation with Weyl anomaly \cite{Cai:2009ua} and for a toy model of Einstein gravity with a GB classically ``entropic'' term that mimics a quantum correction \cite{Cognola:2013fva}.

In the context of gravitational wave detection, quasinormal modes (QNMs) and quasinormal frequencies (QNFs) play an important role \cite{Regge:1957td,Zerilli:1971wd, Kokkotas:1999bd, Nollert:1999ji, Konoplya:2011qq}. Although the detected signal is consistent with Einstein gravity \cite{TheLIGOScientific:2016src}, the large uncertainties in the mass and angular momentum of the ringing black hole leave room for alternative theories of gravity \cite{Konoplya:2016pmh}. The QNMs of a scalar field in higher-dimensional EGB gravity
have been studied in \cite{Iyer:1989rd, Abdalla:2005hu}, where the effect of the coupling constant $\alpha$ was investigated. Furthermore, in Refs. \cite{Konoplya:2017ymp, Gonzalez:2017gwa, Grozdanov:2016fkt, Grozdanov:2016vgg, Gonzalez:2018xrq} the spectrum of QNMs has been shown in theories with higher curvature corrections, such as EGB gravity with a negative cosmological constant, to consist of two distinct branches. One branch has an Einsteinian limit when the GB coupling constant $\alpha$ approaches zero, while the other branch consists of purely imaginary modes, and these modes are qualitatively different from their Einsteinian counterparts. This branch is therefore non-perturbative in $\alpha$ \cite{Gonzalez:2017gwa}. However, calling non-perturbative to this branch could sound inappropriate because it is derived by solving the linearized (perturbative) scalar
equation.

In this work we consider $4D$-EGB black holes with de Sitter (dS) asymptotics and study the propagation of massive scalar fields in such backgrounds, to describe the distinct branches of modes and their decay rates. In particular, we describe the behavior of the perturbative (in $\alpha$) Schwarzschild branch for small and high values of the angular number $\ell$, where we show the existence of anomalous behavior of the decay rate of the QNMs. The anomalous behavior occurs when the longest-lived modes are the ones with higher angular number, and this can occur with massless and massive probe scalar fields. There is a critical mass of the scalar field where the behavior of the decay rate of the QNMs is inverted and can be obtained from the condition $Im(\omega)_{\ell}=Im(\omega)_{\ell+1}$ in the {\it eikonal} limit, that is when $\ell \rightarrow \infty$. The anomalous behavior in the QNFs is possible in asymptotically flat spacetimes, asymptotically dS, and asymptotically AdS spacetimes; however, we observed that the critical mass exists for asymptotically flat spacetimes and for asymptotically dS spacetimes and is not present in asymptotically AdS spacetimes for large and intermediate BHs.

This behavior has been studied for scalar fields \cite{Konoplya:2004wg, Konoplya:2006br,Dolan:2007mj, Tattersall:2018nve,Lagos:2020oek, Aragon:2020tvq,Aragon:2020xtm,Fontana:2020syy,Destounis:2020pjk,Becar:2023jtd,Gonzalez:2022ote,Becar:2023zbl,Alfaro:2024tdr, Becar:2024agj} as well as charged scalar fields \cite{Gonzalez:2022upu,Becar:2022wcj} and fermionic fields \cite{Aragon:2020teq}. 
Also, we describe the perturbative (in $\alpha$) de Sitter branch, which is characterized by purely imaginary QNFs for a massless scalar field that can acquire a real part for a massive scalar field, 
and the non-perturbative (in $\alpha$) de Sitter branch of modes, with purely imaginary QNFs. However, purely imaginary QNFs from both de Sitter branches can combine, resulting in complex frequencies. It is worth mentioning that the QNMs of the test fields in this background was analyzed in \cite{Churilova:2020mif} and the instability of propagation of charged scalar fields in \cite{Liu:2020evp}. Also, the QNFs of Dirac's field was studied in Ref. \cite{Churilova:2020aca}, and it was shown that the real part of the QNFs is considerably increased, while the damping rate usually decreases when the coupling constant increases. 

The manuscript is organized as follows: in Sect. \ref{background} we give a brief review of the $4D$-EGB gravity. In Sect. \ref{QNM}, we study the massive scalar field perturbations. Then, in Sect. \ref{families} we describe the branches of modes and their decay rates in Sect. \ref{Dom}. Finally, our conclusions are given in Sect. \ref{conclusion}.

\section{Einstein-Gauss-Bonnet black hole in four-dimensional de Sitter spacetime}
\label{background}
The Lagrangian of the $D$-dimensional EGB theory with the coupling constant re-scaled by $\alpha \rightarrow\frac{\alpha}{D-4}$, is given by the relation \cite{Fernandes:2020rpa}
\begin{equation}
\mathcal{L}=R-2 \Lambda+\frac{\alpha}{D-4}\mathcal{G}
\, ,
\end{equation}
where $R$ is the Ricci scalar, $\Lambda=-\frac{(D-1)(D-2)}{2 l^2}$ is the cosmological constant in $D$ dimensions and $\mathcal{G}= R^2- 4R_{\mu \nu} R^{\mu \nu} +R_{\mu \nu \rho \sigma} R^{\mu \nu \rho \sigma}$ is the GB term.
The solution for a static and spherically symmetric ansatz in an arbitrary number of dimensions $D\geq 5$, has the form
\begin{equation}
ds^2=-f(r) dt^2+f(r)dr^2+ r^2 d\Omega^2_{D-2} \,,
\end{equation}
where $d\Omega^2_{D-2}$ represents the metric of a $(D-2)$-dimensional hypersphere. Then, following the prescription given in \cite{Glavan:2019inb} and taking the limit $D\rightarrow4$ it is possible to obtain the exact solution representing the $4D$-EGB black hole \cite{Fernandes:2020rpa}
\begin{equation}  \label{metrica}
f(r)=1+\frac{r^2}{2\alpha}\left(1\pm\sqrt{1+4\alpha\left(\frac{2M}{r^3}+\frac{\Lambda}{3}\right)}\right) \,,
\end{equation}
where $M$ denotes the mass of the black hole. It is worth mentioning that similar metrics were previously found in the context of quantum corrections to gravity \cite{Tomozawa:2011gp, Cognola:2013fva, Cai:2009ua}.
Among the two solution branches, we focus on the negative branch, as it recovers the well-known Einsteinian limits-- namely, the asymptotically flat, de Sitter and anti-de Sitter spacetimes-- when the coupling constant $\alpha$ approaches zero.
For instance, rewriting $f(r)$ as
\begin{equation}
f(r)=1-\frac{\frac{2M}{r}+\frac{\Lambda}{3}r^{2}}{\frac{1}{2}\left(1+\sqrt{1+4\alpha\left(\frac{2M}{r^3}+\frac{\Lambda}{3}\right)}\right)}\,,
\end{equation}
it is straightforward to see that as $\alpha \rightarrow 0$ (the GR limit), the solution, for $\Lambda>0$, reduces to the Schwarzschild-de Sitter (SdS) black hole. Additionally, as $r\rightarrow \infty$, the metric asymptotically approaches de Sitter spacetime with an effective positive cosmological constant. In the special case where $\Lambda=0$, the spacetime becomes asymptotically flat.
The central singularity in Eq. (\ref{metrica}) exhibits a repulsive nature, contrasting with its attractive behavior in higher-dimensional spacetimes.
From this point forward, we will work with dimensionless quantities; then the metric function will be as follows
\begin{equation}
f(\tilde{r})=1+\frac{\tilde{r}^2}{2 \tilde{\alpha}} \left(1-\sqrt{\frac{4 \tilde{\alpha}  \left(3 (\tilde{\alpha} +1)+\tilde{\Lambda } \left(\tilde{r}^3-1\right)\right)}{3 \tilde{r}^3}+1}\,\right)\,,
\end{equation}
where $\tilde{r}=\frac{r}{r_{H}}$, $\tilde{\Lambda}=\Lambda r_{H}^{2}$ and $\tilde{\alpha}=\frac{\alpha}{r_{H}^{2}}$. In this case, in the limit $\tilde{\alpha}\rightarrow 0$, the metric function reduces to that of the Schwarzschild-de Sitter black hole:
\begin{equation}
f(\tilde{r})=1-\frac{1}{\tilde{r}}+\frac{\tilde{\Lambda}}{3 \tilde{r}}(1-\tilde{r}^{3})\,.   
\end{equation}

\begin{figure}[h!] 
\begin{center}
\includegraphics[width=0.4\textwidth]{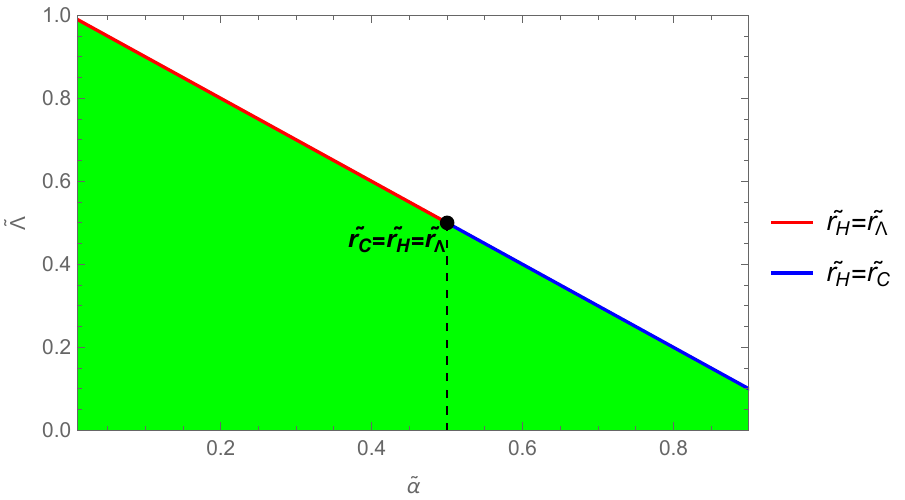}
\end{center}
\caption{The green region represents the range of dimensionless parameters $\tilde{\Lambda}$ and $\tilde{\alpha}$ for which the black hole solution has three distinct horizons: the event horizon, the Cauchy (inner) horizon, and the cosmological horizon. The boundaries of this region correspond to the conditions $\tilde{r}_{H} = \tilde{r}_{C}$ (blue line)
and $\tilde{r}_{\Lambda} = \tilde{r}_{C}$ (red line).
} 
\label{fr3}
\end{figure}

\begin{figure}[h!] 
\begin{center}
\includegraphics[width=0.4\textwidth]{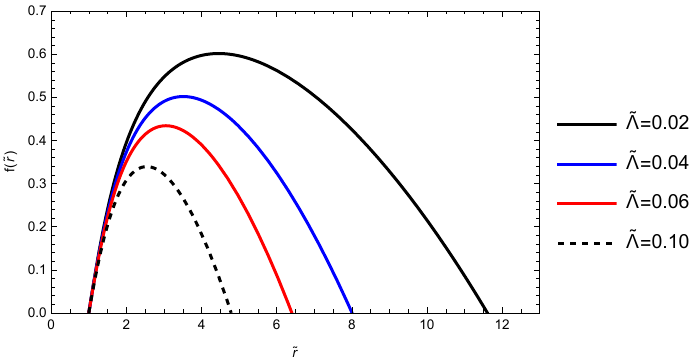}
\includegraphics[width=0.4\textwidth]{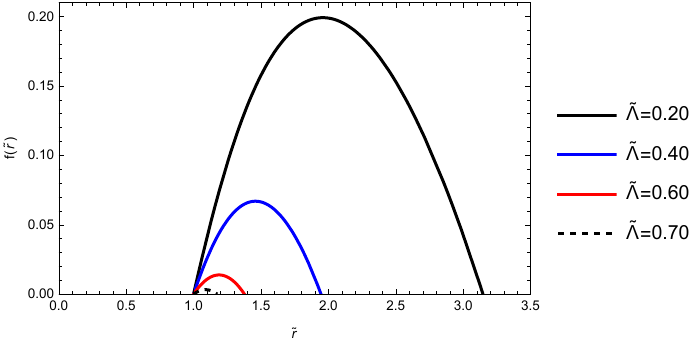}
\end{center}
\caption{The behavior of the metric function $f(\tilde{r})$ as a function of $\tilde{r}$ for different values of the parameter $\tilde{\Lambda}$ with  $\tilde{\alpha}=0.2$.} 
\label{fr1}
\end{figure}

\begin{figure}[h!] 
\begin{center}
\includegraphics[width=0.4\textwidth]{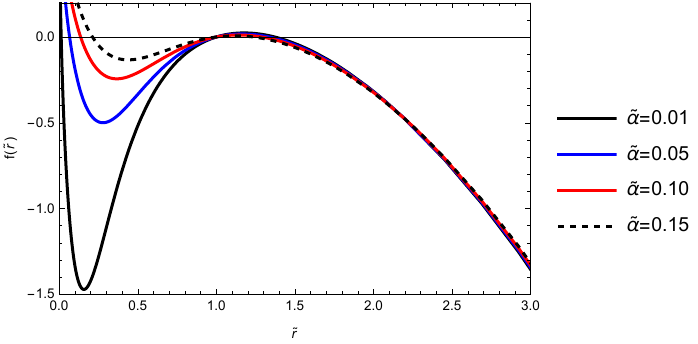}
\includegraphics[width=0.4\textwidth]{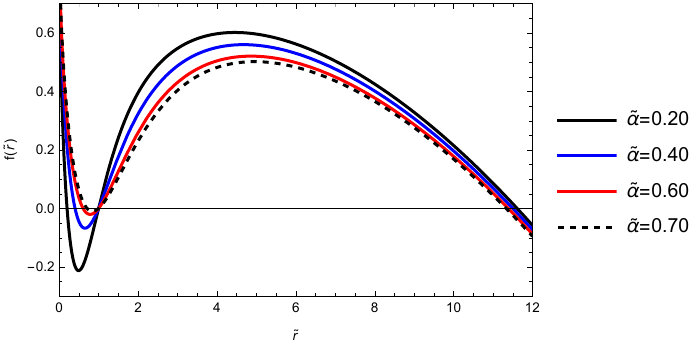}
\end{center}
\caption{The behavior of the metric function $f(\tilde{r})$ as a function of $\tilde{r}$ for different values of the parameter $\tilde{\alpha}$ with  $\tilde{\Lambda}=0.02$ (left panel), and $\tilde{\Lambda}=0.70$ (right panel).} 
\label{fr2}
\end{figure}

Fig. \ref{fr3} shows the parameter space of $\tilde{\Lambda}$ and $\tilde{\alpha}$ for which the 4-dimensional Einstein-Gauss-Bonnet de Sitter black hole has three distinct horizons: the event horizon $\tilde{r}_{H}$, the Cauchy horizon $\tilde{r}_{C}$, and the cosmological horizon $\tilde{r}_{\Lambda}$. In the region above the green area, only the event horizon exists, given by $\tilde{r}_{H}=1$. 
The metric function is shown in Fig. \ref{fr1}, for both small values of $\tilde{\Lambda}$ (left panel) and larger values of $\tilde{\Lambda}$ (right panel), with $\tilde{\alpha}=0.2$. We observe that as ${\tilde{\Lambda}}$ increases, the cosmological horizon $\tilde{r}_{\Lambda}$ decreases. This reduction also decreases the maximum value of the metric function, thereby altering its asymptotic behavior in the far-field region. Furthermore, as $\tilde{\Lambda}$ increases, the event horizon $\tilde{r}_H$ approaches $\tilde{r}_{\Lambda}$. 
On the other hand, the metric function is shown in Fig. \ref{fr2}, for small values of $\tilde{\alpha}$ and $\tilde{\Lambda}=0.02$ (left panel) and for larger values of $\tilde{\alpha}$ and $\tilde{\Lambda}=0.70$ (right panel). We observe that as the GB parameter $\tilde{\alpha}$ increases, the Cauchy horizon approaches the event horizon, and the black hole becomes near-extremal. In addition, the maximum value of the metric function decreases. This indicates that an increase in $\tilde{\alpha}$ significantly affects the near-field spacetime structure. It is important to note that our analysis is based on positive values of $\tilde{\alpha}$, where the existence of Cauchy, event, and cosmological horizons is guaranteed. In contrast, for the Schwarzschild-dS black hole, the Cauchy horizon is always negative and lacks physical significance.


\section{Massive scalar field perturbations}
\label{QNM}

\begin{figure}[h!]
\begin{center}
\includegraphics[width=0.358\textwidth]{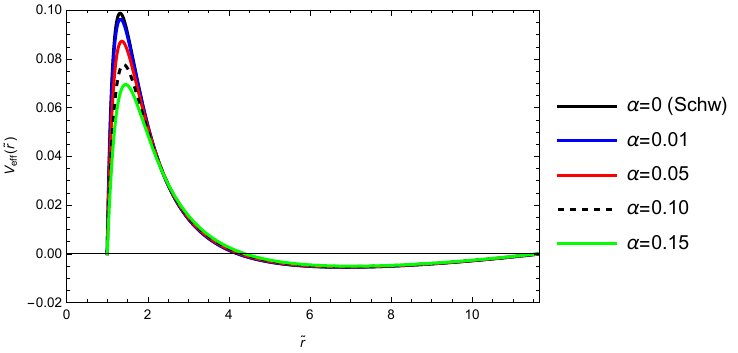}
\includegraphics[width=0.31\textwidth]{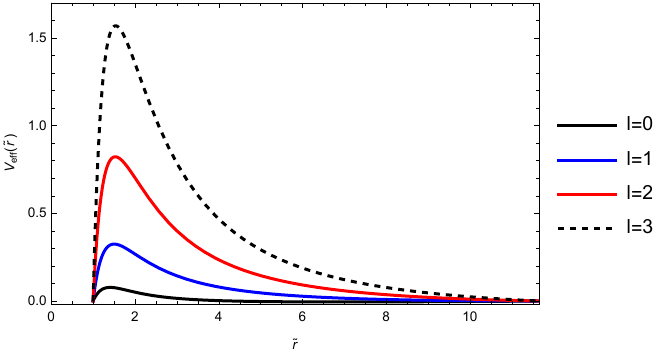}
\includegraphics[width=0.31\textwidth]{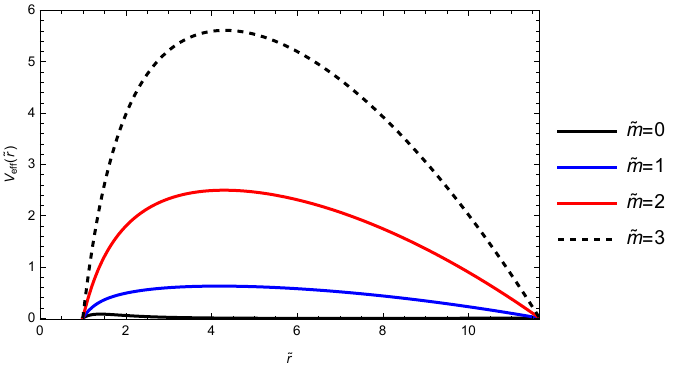}
\end{center}
\caption{The behavior of the effective potential  $V_{eff}$ as a function of $\tilde{r}$ with  $\tilde{\Lambda}=0.02$. 
Left panel for different values of the parameter $\tilde{\alpha}$, with $\tilde{m}=0$, $\ell=0$. Central panel for different values of the parameter $\ell$, with $\tilde{m}=0$, and $\tilde{\alpha}=0.10$, and right panel for different values of the parameter $\tilde{m}$, with $\ell=0$, and $\tilde{\alpha}=0.10$.}  
\label{Potential1}
\end{figure}

The QNMs of scalar field perturbations in the background of  the metric (\ref{metrica}) are determined by the solution to the Klein-Gordon equation
\begin{equation}
\frac{1}{\sqrt{-g}}\partial _{\mu }\left( \sqrt{-g}g^{\mu \nu }\partial_{\nu } \varphi \right) =m^{2}\varphi \,,  \label{KGNM}
\end{equation}%
with appropriate boundary conditions for a black hole geometry. In the above expression $m$ denotes the mass of the scalar field $\varphi $. Now, by means of the following ansatz
\begin{equation}
\varphi =e^{-i\omega t} R(r) Y(\Omega) \,,\label{wave}
\end{equation}%
the Klein-Gordon equation reduces to
\begin{equation}
f(r)R''(r)+\left(  f'(r)+2\frac{f(r)}{r} \right)R'(r)+\left(\frac{\omega^2}{f(r)}-\frac{\ell (\ell+1) }{r^2}-m^{2}\right) R(r)=0\,, \label{radial}
\end{equation}%
where $\ell=0,1,2,...$ represents the azimuthal quantum number and the prime denotes the derivative with respect to $r$.
Now, defining $R(r)=\frac{F(r)}{r}$
and by using the tortoise coordinate $r^*$ defined by
$dr^*=\frac{dr}{f(r)}$, the Klein-Gordon equation can be rewritten as a one-dimensional Schr\"{o}dinger equation
\begin{equation}\label{ggg}
\frac{d^{2}F(r^*)}{dr^{*2}}-V_{eff}(r)F(r^*)=-\omega^{2}F(r^*)\,,
\end{equation}
where the effective potential $V_{eff}(r)$, parametrically thought as $V_{eff}(r^*)$, is given by
\begin{equation}\label{pot}
V_{eff}(r)=f(r) \left(\frac{f'(r)}{r}+\frac{\ell (\ell+1)}{r^2} + m^2 \right)~.
\end{equation}
In terms of dimensionless quantities, Eq. (\ref{ggg}) can be written as
\begin{equation}
\frac{d^{2}F(\tilde{r}^*)}{d\tilde{r}^{*2}}-V_{eff}(\tilde{r})F(\tilde{r}^*)=-\tilde{\omega}^{2}F(\tilde{r}^*)\,,
\end{equation}
where 
\begin{equation}
V_{eff}(\tilde{r})=\frac{f(\tilde{r})}{\tilde{r}^2} \left(\kappa^{2} + \tilde{m}^2 \tilde{r}^2+f^\prime(\tilde{r})\tilde{r}\right)\,,
\end{equation} 
with $\tilde{\omega} = \omega r_h$, and $\tilde{m}= m r_h$.  The effective potential is shown in Fig. \ref{Potential1}. Note that the maximum value of the effective potential corresponds to the Schwarzschild-dS black hole,  and this maximum increases as the parameter $\tilde{\alpha}$ decreases (left panel), and as either the parameter $\ell$ (central panel) or $\tilde{m}$ (right panel) increases. Also, observe that there is a region where the effective potential takes negative values (left panel).
This behavior for $\ell =0$, has been observed in other spacetimes, where the effective potential is negative only for $\ell=0$, while it remains positive for $\ell \neq 0$. In these cases, the propagation of massive scalar fields is stable. However, when the field is charged, the propagation becomes unstable for $\ell=0$ in this spacetime, as discussed in Ref. \cite{Liu:2020evp}.

\section{Quasinormal spectrum}
\label{families}

In this section we show that the introduction of the Gauss-Bonnet coupling constant $\tilde{\alpha}$ gives rise to three branches of QNMs: the perturbative (in $\alpha$) Schwarzschild branch, the perturbative (in $\alpha$) dS branch, 
and a non-perturbative  (in $\alpha$) dS branch.

\begin{figure}[H]
\begin{center}
\includegraphics[width=0.6\textwidth]{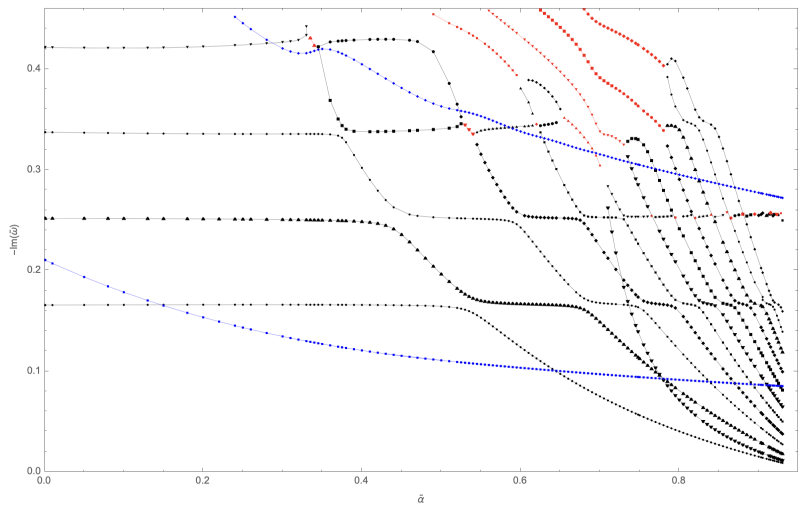}
\end{center}
\caption{The behavior of $-Im(\tilde{\omega})$ as a function of the rescaled Gauss-Bonnet coupling constant $\tilde{\alpha}$ for a massless scalar field with $\ell=0$  and $\tilde{\Lambda}=0.02$.
The blue lines represent the Schwarzschild branch, characterized by complex frequencies with nonzero real parts. The black lines corresponds to the dS branches, associated with purely imaginary frequencies. In certain regions, two purely imaginary frequencies coalesce into a complex frequency, depicted by the red segments. Conversely, a complex frequencies (red segment) can bifurcate into two purely imaginary frequencies. Additionally, for $m=0$ and $\ell=0$, there exist a zero mode with $\omega=0$, which is not shown in the figure.}
\label{Dominanceell0}
\end{figure}

To distinguish between the different branches of modes, we plot in Fig. \ref{Dominanceell0} the behavior of $-Im(\tilde{\omega})$ as a function of the rescaled Gauss-Bonnet coupling constant $\tilde{\alpha}$, for $\ell=0$ and $\tilde{\Lambda}=0.02$. 
The blue lines in the figure correspond to a branch comprising complex frequencies with non-zero real parts.
This branch smoothly approaches the Schwarzschild limit as $\tilde{\alpha} \rightarrow 0$, indicating its perturbative nature in $\tilde{\alpha}$ and its correspondence to the Einsteinian regime, namely the perturbative Schwarzschild branch. In contrast, the black lines correspond to purely imaginary frequencies, delineating two distinct dS branches, which are absent for $\Lambda=0$. One of these is perturbative in $\tilde{\alpha}$, showing a well-defined Einsteinian limit as $\tilde{\alpha} \rightarrow 0$. The other is non-perturbative, lacking an Einsteinian limit due to the divergence of its damping rates as $\tilde{\alpha}$ decreases; these modes are absent at $\tilde{\alpha}=0$ and become longest-lived for larger values of $\tilde{\alpha}$. To distinguish between these two branches of the dS, we note that the perturbative dS frequencies vary slowly with $\tilde{\alpha}$, corresponding to the nearly horizontal segment of the black lines in the figure. This behavior is appreciable for low overtones; however, it becomes less clear for high overtones. In contrast, non-perturbative dS frequencies exhibit significant variation as $\tilde{\alpha}$ decreases, reflecting their non-perturbative character. However, it is important to note that the purely imaginary frequencies from both dS branches can merge at the red segments of the lines, representing complex frequencies, making them indistinguishable from each other. It is worth mentioning that the behavior of the branches is the same for the case $\ell=1$, see Fig. 
\ref{Dominance1}.


\subsection{Perturbative Schwarzschild branch}

We employ two different numerical methods to compute the QNFs. One of them is the pseudospectral Chebyshev technique, which is used for small values of $\ell$, and the other method is the 
sixth-order Wentzel-Kramers-Brillouin (WKB) method with Padé approximants, 
which is applied for large values of $\ell$. In order to provide analytical insights into the behavior of QNFs in the eikonal limit ($\ell \rightarrow \infty$), and to
determine the critical scalar field mass 
we use the third order WKB approximation. In Appendix \ref{Accuracy} we show the accuracy of the numerical technique used.

\subsubsection{Small values of the angular number}

To solve the differential equation (\ref{radial}) numerically and compute the QNFs, we employ the pseudospectral Chebyshev method, as described, for example, in \cite{Boyd}. First, it is convenient to perform a change of variables to map the radial coordinate to the interval $[0,1]$. Specifically, we define a new variable $y=(r-r_H)/(r_{\Lambda}-r_H)$, so that the event horizon corresponds to $y=0$, and the cosmological horizon corresponds to $y=1$. With this change of variables, the radial equation (\ref{radial}) becomes
\begin{eqnarray} \label{rad}
&&\nonumber f(y) R''(y) + \left( \frac{2 \left(    \tilde{r}_{\Lambda}- \tilde{r}_H  \right) f(y)}{1+\left( \tilde{r}_{\Lambda}-\tilde{r}_H \right) y } + f'(y) \right) R'(y)\\
&&+ \left( \tilde{r}_{\Lambda}-\tilde{r}_H  \right)^2 \left( \frac{\tilde{\omega}^2}{f(y)}- \frac{ \ell(\ell+1)}{\left( 1 + \left( \tilde{r}_{\Lambda}-\tilde{r}_H \right)y \right)^2} -\tilde{m}^2  \right) R(y)=0\,.
\end{eqnarray}
In the vicinity of the horizon $(y \rightarrow 0)$, the function $R(y)$ behaves as
\begin{equation}
R(y)=C_1 e^{-\frac{i \tilde{\omega} \left( \tilde{r}_{\Lambda}-\tilde{r}_H \right)}{f'(0)} \ln{y}}+C_2 e^{\frac{i \tilde{\omega} \left( \tilde{r}_{\Lambda}-\tilde{r}_H \right)}{f'(0)} \ln{y}} \,.
\end{equation}
Here, the first term represents an ingoing wave, while the second term represents an outgoing wave near the black hole horizon.
By imposing the condition of only ingoing waves at the horizon, we set $C_2=0$. On the other hand, near the cosmological horizon, the function $R(y)$ behaves as
\begin{equation}
R(y)= D_1 e^{-\frac{i \tilde{\omega} \left( \tilde{r}_{\Lambda}-\tilde{r}_H \right)}{f'(1)} \ln{(1-y)}}+D_2 e^{\frac{i \tilde{\omega} \left( \tilde{r}_{\Lambda}-\tilde{r}_H \right)}{f'(1)} \ln{(1-y)}}  \,.
\end{equation}
Here, the first term represents an outgoing wave, while the second term represents an ingoing wave near the cosmological horizon. By imposing the condition of only ingoing waves at the cosmological horizon, we set $D_1=0$.
Considering the behavior of the scalar field at both the event and cosmological horizons, we define the following ansatz
\begin{equation}
R(y)= e^{-\frac{i \tilde{\omega} \left( \tilde{r}_{\Lambda}-\tilde{r}_H \right)}{f'(0)} \ln{y}} e^{\frac{i \tilde{\omega}  \left( \tilde{r}_{\Lambda}-\tilde{r}_H \right)  }{f'(1)} \ln{(1-y)}} F(y) \,.
\end{equation}
By substituting the ansatz for $R(y)$ into Eq. (\ref{rad}), a differential equation for the function $F(y)$ is derived. The solution for the function $F(y)$ is assumed to take the form of a finite linear combination of Chebyshev polynomials, which is then substituted into the differential equation for $F(y)$. The interval $[0,1]$ is discretized at the Chebyshev collocation points, and the differential equation is evaluated at each collocation point. This results in a system of algebraic equations, which is identified as a generalized eigenvalue problem. The eigenvalue problem is then solved numerically to obtain the QNFs. 

\begin{figure}[h!]
\begin{center}
\includegraphics[width=0.35\textwidth]{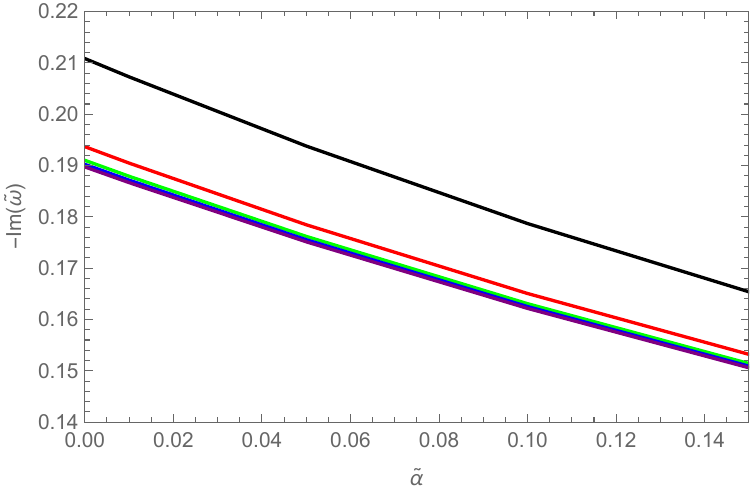}
\includegraphics[width=0.4\textwidth]{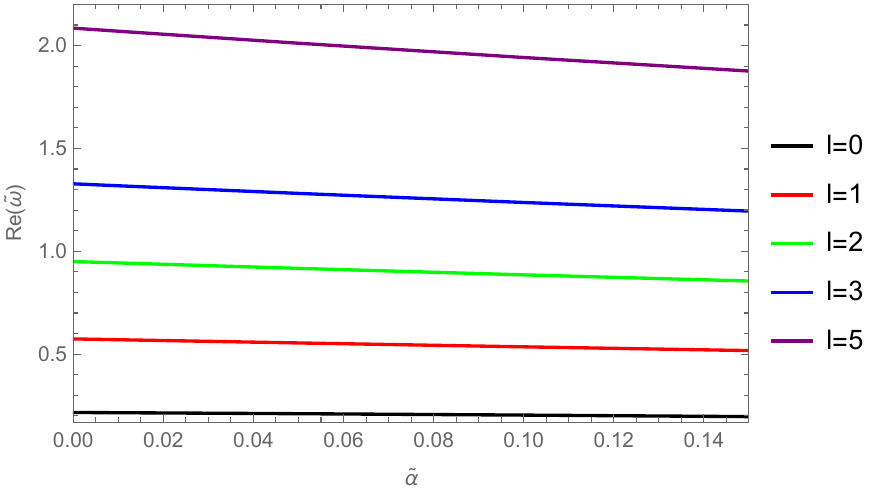}
\end{center}
\caption{The behavior of $-Im(\tilde{\omega})$ (left panel) and $Re(\tilde{\omega})$ (right panel) for the fundamental modes ($n=0$) and angular momentum $\ell=0,1,2,3,5$ as function of the parameter $\tilde{\alpha}$ with $\tilde{m}=0$, and  $\tilde{\Lambda}=0.02$. The QNFs have been computed using the pseudospectral Chebyshev method.}  
\label{ReIm1}
\end{figure}

\begin{figure}[h!]
\begin{center}
\includegraphics[width=0.35\textwidth]{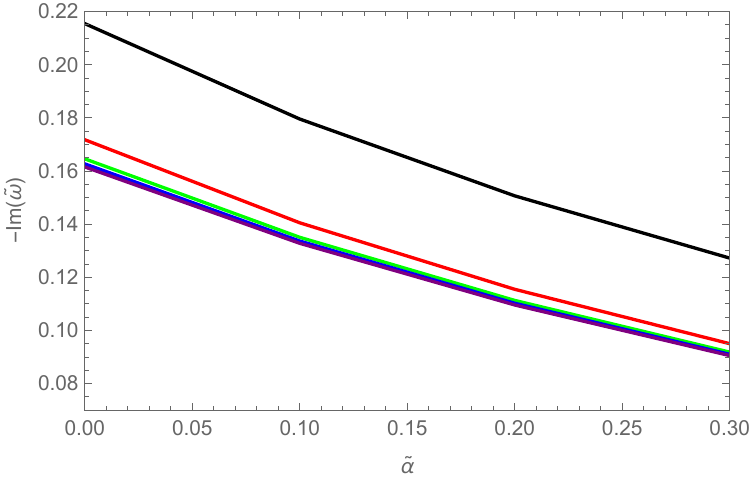}
\includegraphics[width=0.4\textwidth]{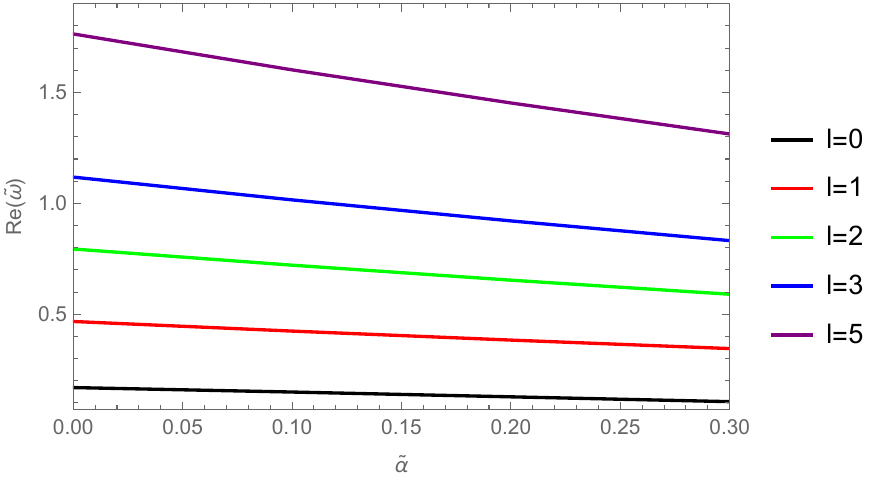}
\end{center}
\caption{The behavior of $-Im(\tilde{\omega})$ (left panel) and $Re(\tilde{\omega})$ (right panel) for the fundamental modes ($n=0$) and angular momentum $\ell=0,1,2,3,5$ as function of the parameter $\tilde{\alpha}$ with $\tilde{m}=0$, and $\tilde{\Lambda}=0.2$. The QNFs have been computed using the pseudospectral Chebyshev method.}  
\label{ReIm2}
\end{figure}

In Figs. \ref{ReIm1} and \ref{ReIm2}, we show the behavior of QNFs as a function of the coupling constant $\tilde{\alpha}$ for small values of the angular number $\ell$, with $\tilde{\Lambda}=0.02$, and 0.2, respectively. We observe that the absolute value of the imaginary part (left panels) increases, while the real part (right panels) decreases as the coupling constant $\tilde{\alpha}$ increases. This indicates that the modes of the perturbative Schwarzschild branch have a larger decay rate and a lower oscillation frequency compared to the SdS background ($\tilde{\alpha}=0$). Also, it is noteworthy that for a massless scalar field, the longest-lived modes correspond to those with higher angular number, a phenomenon known as the anomalous decay rate of the QNMs.
\newpage

\begin{figure}[h!]
\begin{center}
\includegraphics[width=0.345\textwidth]{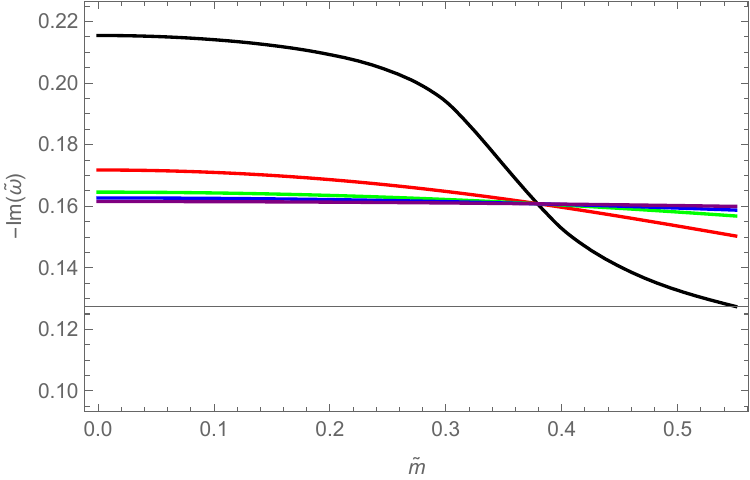}
\includegraphics[width=0.40\textwidth]{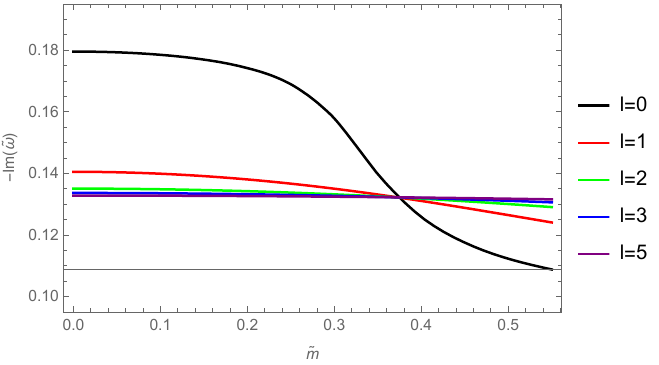}
\end{center}
\caption{The behavior of $-Im(\tilde{\omega})$ for the fundamental modes ($n=0$) and angular momentum $\ell=0,1,2,3,5$ as function of the scalar field mass ($\tilde{m})$ with $\tilde{\Lambda}=0.2$. Left panel for $\tilde{\alpha}=0$, and right panel for $\tilde{\alpha}=0.1$. The QNFs have been computed using the pseudospectral Chebyshev method.}  
\label{anomalous02ellsmall}
\end{figure}

Furthermore, the effect of the parameter $\tilde{m}$ on $Im(\tilde{\omega})$ is shown in Fig. \ref{anomalous02ellsmall}. It can be observed that the anomalous decay rate of the QNMs for the massless scalar field undergoes an inversion at a specific value of the field mass, called the critical scalar field mass ($\tilde{m}_{c}$). Beyond this critical value, i.e., for $\tilde{m}>\tilde{m}_{c}$, the longest-lived modes correspond to those with smaller angular numbers. It is worth mentioning that the values of the parameters ($\tilde{\alpha}$, $\tilde{\Lambda}$) in Fig. \ref{anomalous02ellsmall} are such that the modes of the perturbative Schwarzschild branch have the smallest decay rates than the other branches. When this is not the case, the $\ell=0$ modes exhibit a different behavior. Also, note that the critical scalar field mass is lower than that of the SdS background for the values of the parameters considered. 

\subsubsection{High values of the angular number}
\label{WKBJ}

We employ the method based on the WKB approximation, originally proposed by Mashhoon \cite{Mashhoon} and further developed by Schutz and Iyer \cite{Schutz:1985km}. Iyer and Will computed the third order correction \cite{Iyer:1986np}, which was later extended to the sixth order \cite{Konoplya:2003ii}, and more recently to the thirteenth order \cite{Matyjasek:2017psv}, see also \cite{Konoplya:2019hlu}.\\
This method has been effectively applied to determine the QNFs for both asymptotically flat and asymptotically de Sitter black holes. The WKB method is particularly applicable for effective potentials that resemble potential barriers, which approach constant values at both the horizon and spatial infinity \cite{Konoplya:2011qq}. However, it is important to note that this approach can only provide modes corresponding to the Schwarzschild branch. The QNMs are determined by the behavior of the effective potential near its maximum value $V(r^*_{max})$. The Taylor series expansion of the potential around its maximum is given by the following expression
\begin{equation}
V(r^*)= V(r^*_{max})+ \sum_{i=2}^{i=\infty} \frac{V^{(i)}}{i!} (r^*-r^*_{max})^{i} \,,
\end{equation}
where
\begin{equation}
V^{(i)}= \frac{d^{i}}{d r^{*i}}V(r^*)|_{r^*=r^*_{max}}\,,
\label{eq:derivadas}
\end{equation}
corresponds to the $i$-th derivative of the potential with respect to $r^*$, evaluated at the position of the maximum of the potential, $r^*_{max}$. Using the WKB approximation up to third order beyond the eikonal limit, the QNFs are given by the following expression \cite{Iyer:1986np}
\begin{eqnarray}
\omega^2 &=& V(r^*_{max})  -2 i U \,,
\end{eqnarray}
where
\begin{eqnarray}
\notag U &=&  N\sqrt{-V^{(2)}/2}+\frac{i}{64} \left( -\frac{1}{9}\frac{V^{(3)2}}{V^{(2)2}} (7+60N^2)+\frac{V^{(4)}}{V^{(2)}}(1+4 N^2) \right) +\frac{N}{2^{3/2} 288} \Bigg( \frac{5}{24} \frac{V^{(3)4}}{(-V^{(2)})^{9/2}} (77+188N^2) + \\
\notag && \frac{3}{4} \frac{V^{(3)2} V^{(4)}}{(-V^{(2)})^{7/2}}(51+100N^2) +\frac{1}{8} \frac{V^{(4)2}}{(-
V^{(2)})^{5/2}}(67+68 N^2)+\frac{V^{(3)}V^{(5)}}{(-V^{(2)})^{5/2}}(19+28N^2)+\frac{V^{(6)}}{(-V^{(2)})^{3/2}} (5+4N^2)  \Bigg)\,,
\end{eqnarray}
and $N=n+1/2$, with $n=0,1,2,\dots$, is the overtone number.
The imaginary and real part of the QNFs can be written as
\begin{eqnarray}
\label{im} \omega_I^2 &=& - (Im(U)+V/2)+\sqrt{(Im(U)+V/2)^2+Re(U)^2} \,, \\
\omega_R^2 &=& -Re(U)^2 / \omega_I^2 \,,
\end{eqnarray}
respectively, where $Re(U)$ denotes the real part of $U$, and $Im(U)$ represents its imaginary part. In Appendix \ref{wkba}, the full expression for the effective potential evaluated at $r^{*}_{max}$ is provided, along with the expansion of the QNFs for large values of 
$\ell$. It is important to note that the critical mass of the scalar field, denoted as $\tilde{m}_c$, is defined as the mass value at which $\tilde{\omega}_{2}$ is nullified. However, since the complete expression for the critical mass is quite lengthy, we will provide an approximate fourth order equation, which is valid for small values of the rescaled GB parameter $\tilde{\alpha}$, see Appendix \ref{wkba}.

\begin{figure}[H]
\begin{center}
\includegraphics[width=0.40\textwidth]{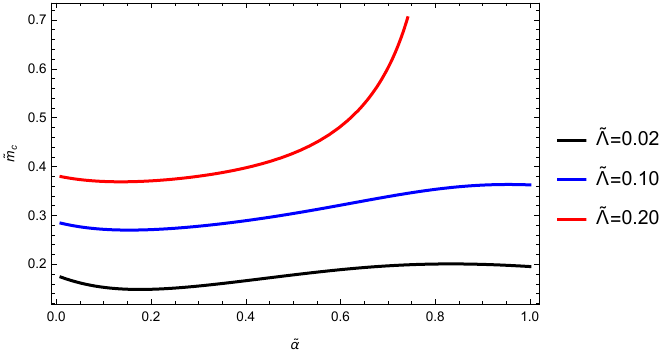}
\includegraphics[width=0.40\textwidth]{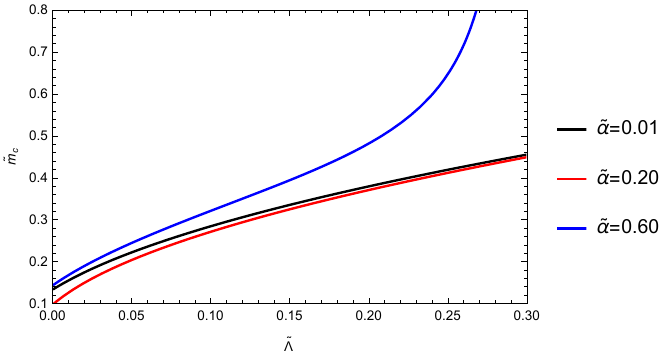}
\end{center}
\caption{The behavior of the exact critical scalar field mass $\tilde{m}_{c}$  as a function of the rescaled GB coupling constant $\tilde{\alpha}$  for $\tilde{\Lambda}=0.02, 0.10, 0.20$ (left panel), and as a function of $\tilde{\Lambda}$ for $\tilde{\alpha}=0.01, 0.20, 0.60$ (right panel).}  
\label{mcex1}
\end{figure}

We show the behavior of the exact critical scalar field mass $\tilde{m}_{c}$ as a function of $\tilde{\alpha}$ in Fig. \ref{mcex1} (left panel). As $\tilde{\alpha}$ increases, the critical scalar field mass initially decreases until it reaches a minimum value, after which it increases with further increases in $\tilde{\alpha}$. Furthermore, for a fixed value of $\tilde{\alpha}$, the critical scalar field mass increases as $\tilde{\Lambda}$ increases. In Fig. \ref{mcex1} (right panel), we show the behavior of $\tilde{m}_{c}$ as a function of $\tilde{\Lambda}$. We observe that $\tilde{m}_{c}$ increases as $\tilde{\Lambda}$ increases. Furthermore, for a fixed value of $\tilde{\Lambda}$, the critical mass initially decreases with increasing $\tilde{\alpha}$, and then starts to increase.

\begin{figure}[H]
\begin{center}
\includegraphics[width=0.42\textwidth]{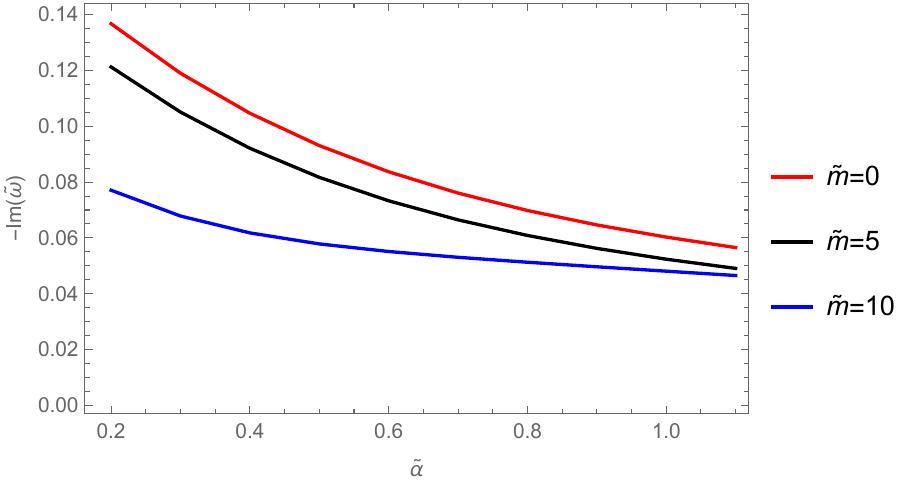}
\includegraphics[width=0.40\textwidth]{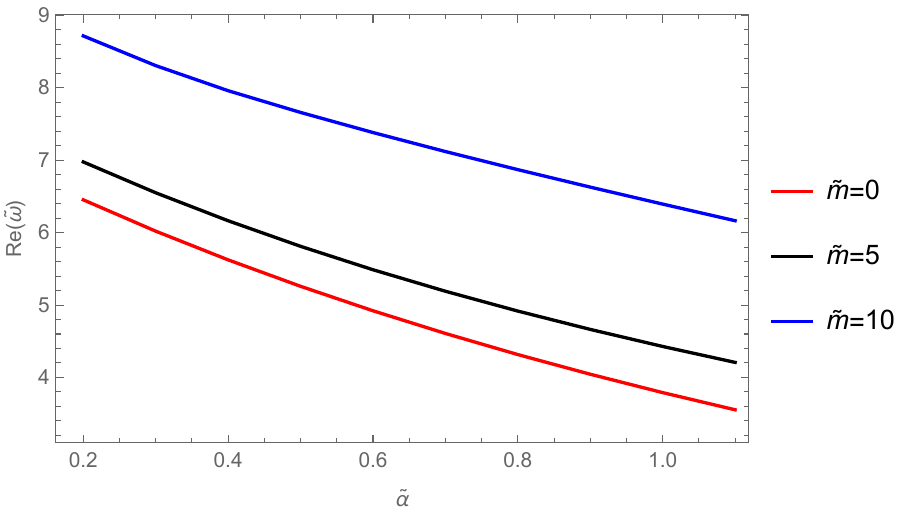}
\end{center}
\caption{The behavior of$-Im(\tilde{\omega})$ (left panel), and $Re(\tilde{\omega})$ (right panel) for the fundamental mode ($n=0$) and scalar field mass $\tilde{m}=0,5.0,10$ as a function of $\tilde{\alpha}$ with $\tilde{\Lambda}= 4/100$, and $\ell=20$.
The QNFs have been computed using the sixth-order WKB method with Padé approximants.}  
\label{mcex2}
\end{figure}

In Fig. \ref{mcex2}, we show the behavior of $-Im(\tilde{\omega})$ and $Re(\tilde{\omega})$ as functions of $\tilde{\alpha}$. We observe that both quantities decrease as $\tilde{\alpha}$ increases. Also, for a fixed value of $\tilde{\alpha}$, $-Im(\tilde{\omega})$ decreases and the oscillation frequency increases as $\tilde{m}$ increases.

\begin{figure}[h!]
\begin{center}
\includegraphics[width=0.30\textwidth]{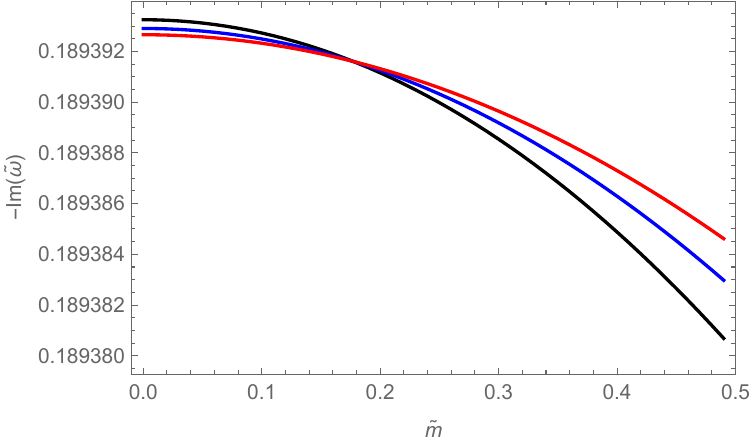}
\includegraphics[width=0.30\textwidth]{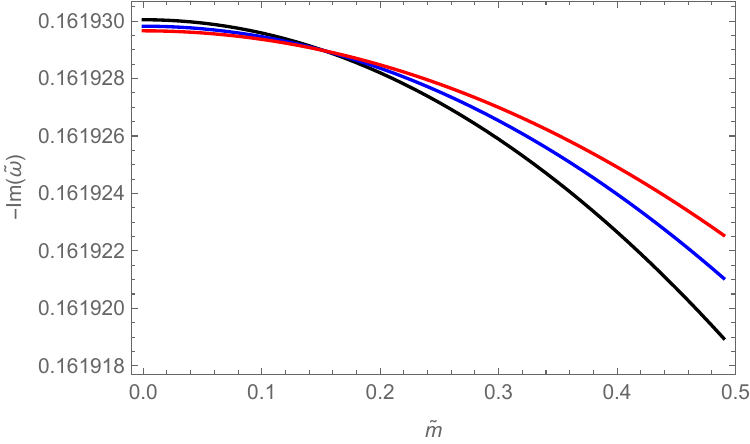}
\includegraphics[width=0.30\textwidth]{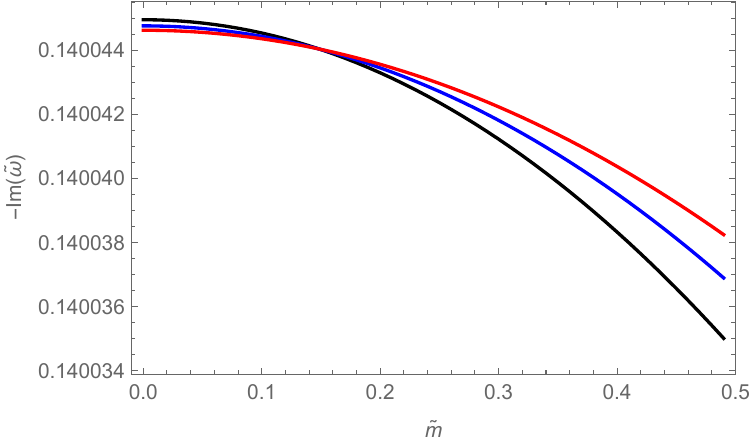}
\end{center}
\caption{The behavior of $-Im(\tilde{\omega})$ as a function of the scalar field mass $\tilde{m}$, computed using the sixth-order WKB method with Padé approximants, is shown for different values of the angular number $\ell=80$ (black line), $\ell=90$ (blue line), and $\ell=100$ (red line). The results are presented for  $\tilde{\Lambda}=0.02$ and three values of the GB parameter: $\tilde{\alpha}=0$ (left panel), $\tilde{\alpha}=0.10$ (central panel), and $\tilde{\alpha}=0.20$ (right panel). The corresponding values of the critical mass, as calculated using Eq. (\ref{mcritical}), are $\tilde{m}_c\approx 0.177$, $\tilde{m}_c\approx 0.153$ and $\tilde{m}_c\approx 0.148$, respectively.}  
\label{ab1}
\end{figure}

\begin{figure}[h!]
\begin{center}
\includegraphics[width=0.30\textwidth]{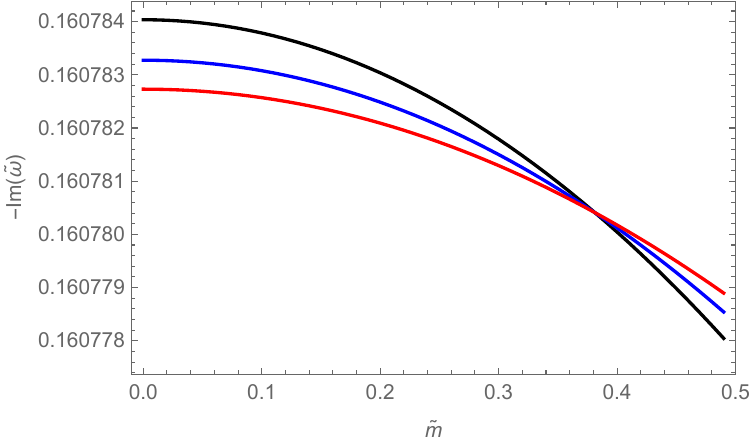}
\includegraphics[width=0.30\textwidth]{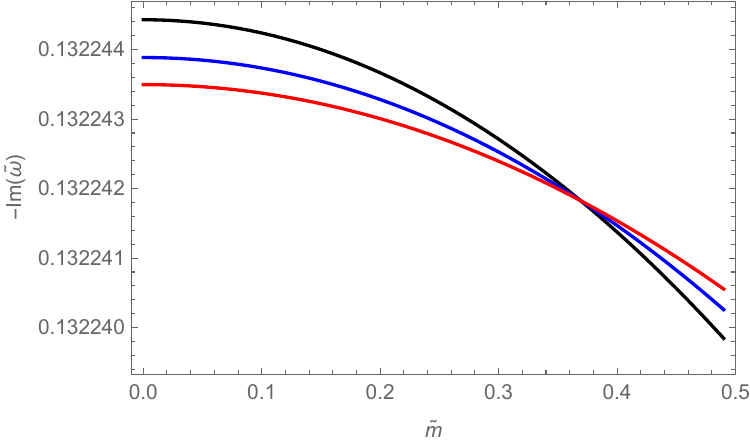}
\includegraphics[width=0.30\textwidth]{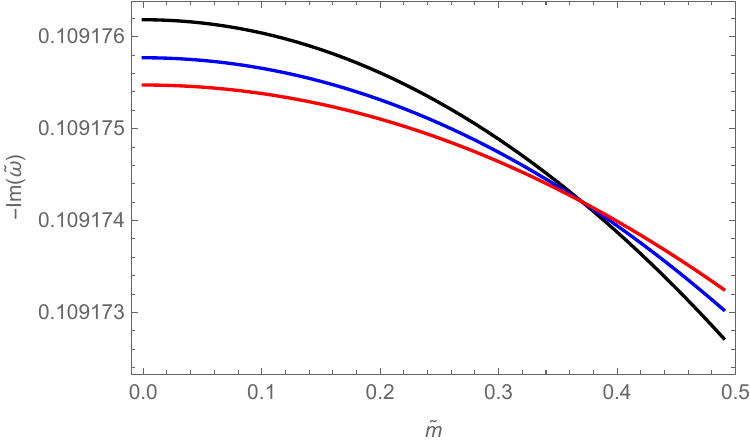}
\end{center}
\caption{The behavior of $-Im(\tilde{\omega})$ as a function of the scalar field mass $\tilde{m}$, computed using the sixth-order WKB method with Padé approximants, is shown for different values of the angular number $\ell=80$ (black line), $\ell=90$ (blue line), and $\ell=100$ (red line). The results are presented for  $\tilde{\Lambda}=0.2$, and three values of the GB parameter: $\tilde{\alpha}=0$ (left panel), $\tilde{\alpha}=0.10$ (central panel), $\tilde{\alpha}=0.20$ (right panel). The corresponding values of the critical mass, as calculated using Eq. (\ref{mcritical}), are $\tilde{m}_c\approx 0.381$, $\tilde{m}_c\approx 0.369$ and $\tilde{m}_c\approx 0.370$, respectively.
}  
\label{ab2}
\end{figure}

To illustrate the anomalous behavior, we plot in Figs. \ref{ab1} and \ref{ab2} the behavior of $-Im(\tilde{\omega})$ as a function of $\tilde{m}$, using the sixth-order WKB method for $\tilde{\Lambda}=0.02$ and $\tilde{\Lambda}=0.2$, respectively. We observe an anomalous decay rate for: $\tilde{m}<\tilde{m}_c$, the longest-lived modes correspond to the highest angular number $\ell$, while for $\tilde{m}>\tilde{m}_c$, the longest-lived modes correspond to the lowest angular number. Furthermore, as the parameter $\tilde{\alpha}$ increases, the behavior observed for $\tilde{m}_c$ is consistent with the trends shown in Fig. \ref{mcex1}.

\subsection{Perturbative de Sitter branch}
\begin{figure}[h]
\begin{center}
\includegraphics[width=0.4\textwidth]{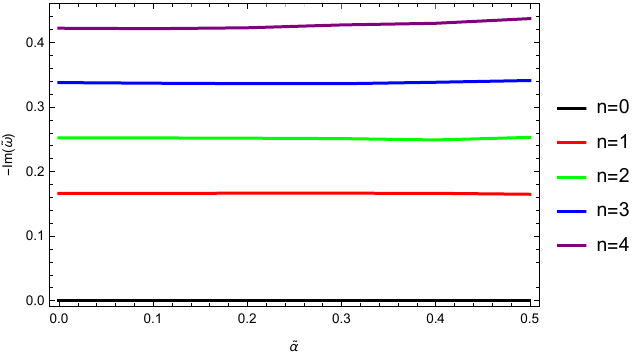}
\includegraphics[width=0.4\textwidth]{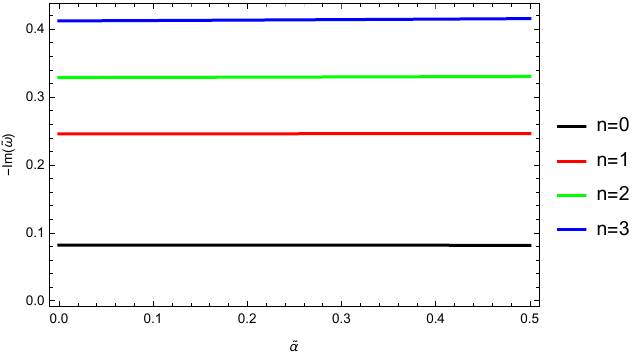}
\end{center}
\caption{The behavior of $-Im(\tilde{\omega})$ for the perturbative dS modes as a function of $\tilde{\alpha}$ is shown for a massless scalar field, with $\tilde{\Lambda}=0.02$, for $\ell=0$ (left panel) and $\ell=1$ (right panel). The QNFs  have been obtained via the pseudospectral Chebyshev method, using a number of Chebyshev polynomials in the range $155$-$160$, with an accuracy of up to ten  decimal places.}
\label{dSm0l01A}
\end{figure}

The QNFs of a pure de Sitter spacetime are given by \cite{Du:2004jt}
\begin{equation}\label{deS}
\omega_{pure-dS} = - i \sqrt{\frac{\Lambda}{3}} \left( 2n_{dS} + \ell + 3/2 \pm \sqrt{\frac{9}{4}-3 \frac{m^2}{\Lambda}} \right)\,,
\end{equation}
where $\Lambda$
is the positive cosmological constant of pure de Sitter spacetime, $n_{dS}= 0, 1, 2, ...$ is the overtone number, and $m$ is the scalar field mass. Note that for a massless scalar field, $\omega_{pure-dS}$ is purely imaginary. The QNMs for scalar perturbations in the background of $4D$--EGB black holes exhibit a perturbative in $\alpha$ de Sitter branch of modes, characterized by purely imaginary frequencies for a massless scalar field, as shown in Fig. \ref{dSm0l01A}. These frequencies are also negative, which means that the propagation of the scalar field is stable in this background. Also, it is important to note that increasing the parameter $\tilde{\alpha}$ does not have a significant effect on the frequencies. These QNFs resemble those of the pure de Sitter spacetime (\ref{deS}), with an effective cosmological constant $\Lambda_{eff}=-\frac{3}{2 \tilde{\alpha}} \left(1 - \sqrt{1 + \frac{4 \tilde{\alpha} \tilde{\Lambda}}{3}}\right)$ and $m=\tilde{m}/r_h$. 
Here, the zero mode with $\tilde{\omega}_{dS}=0$ for $\tilde{m}=0$, $n_{dS}=0$ and $\ell=0$ is treated as a dS mode, as discussed in Ref. \cite{Becar:2023jtd}. It is also important to note that the purely imaginary QNFs of asymptotically de Sitter black holes cannot be computed using the standard WKB method. Therefore, we used the pseudospectral Chebyshev method.

\begin{figure}[h]
\begin{center}
\includegraphics[width=0.45\textwidth]{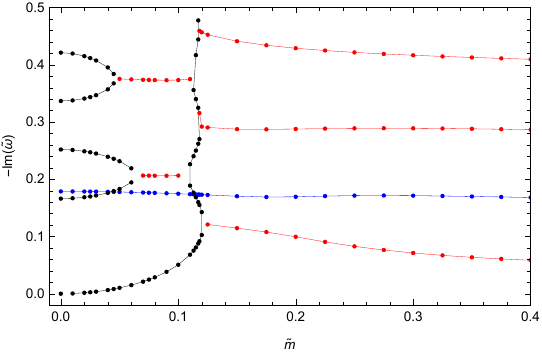}
\includegraphics[width=0.45\textwidth]{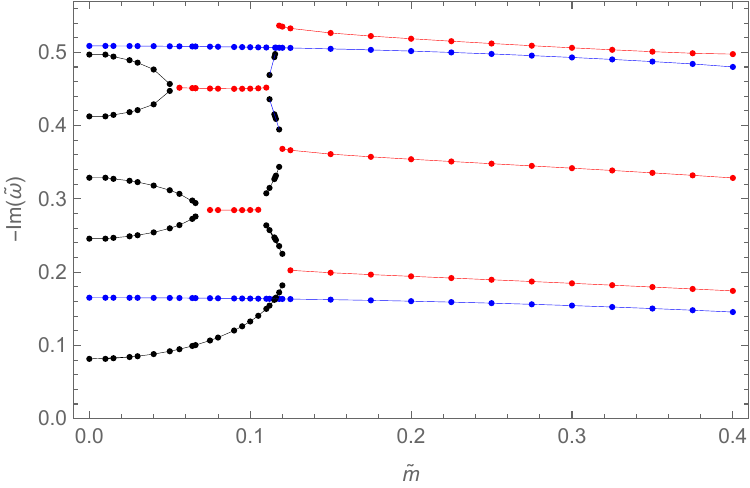}
\end{center}
\caption{The behavior of $-Im(\tilde{\omega})$ for the perturbative dS modes, represented by the black and red points, as a function of $\tilde{m}$ is shown for $\tilde{\Lambda}=0.02$ and $\tilde{\alpha}=0.10$. In the plot, the black points correspond to purely imaginary QNFs, while the red points correspond to the complex QNFs with a non-zero real part. For completeness, the figure also includes the QNFs of the perturbative Schwarzschild branch, represented by the blue points. The left panel shows the results for $\ell=0$, and right panel corresponds to $\ell=1$. The QNFs have been obtained via the pseudospectral Chebyshev method using a number of Chebyshev polynomials in the range $155$-$160$ and an accuracy of ten  decimal places.}
\label{DominancemL002}
\end{figure}

\begin{figure}[h]
\begin{center}
\includegraphics[width=0.45\textwidth]{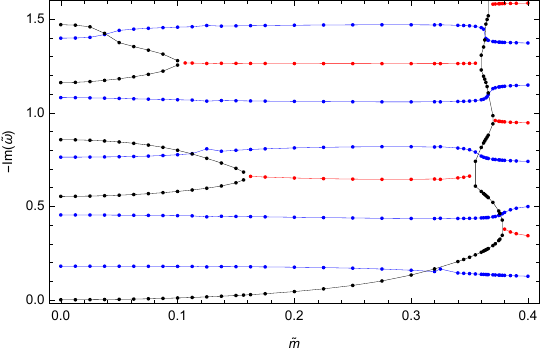}
\includegraphics[width=0.45\textwidth]{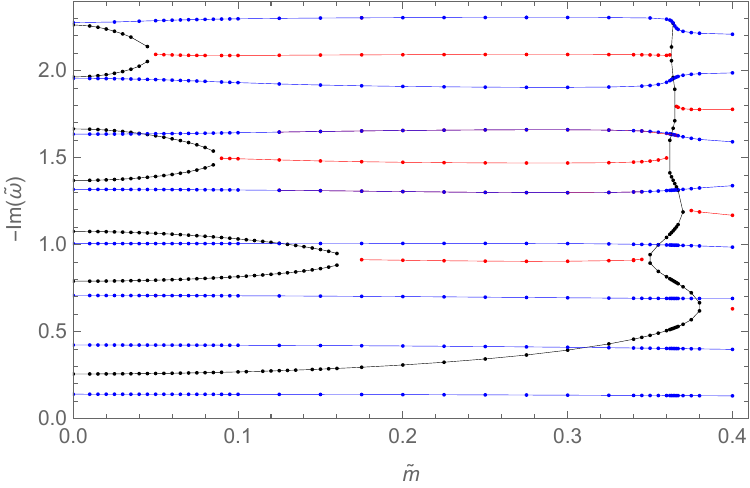}
\end{center}
\caption{The behavior of $-Im(\tilde{\omega})$ for the perturbative dS modes, represented by the black and red points, as a function of $\tilde{m}$ is shown for   $\tilde{\Lambda}=0.2$ and $\tilde{\alpha}=0.10$. In the plot, the black points correspond to purely imaginary QNFs, while the red points correspond to the complex QNFs with a non-zero real part. For completeness, the figure also includes the QNFs of the perturbative Schwarzschild branch, represented by the blue points. The left panel shows the results for $\ell=0$, and the right panel for $\ell=1$. The QNFs have been obtained via the pseudospectral Chebyshev method using a number of Chebyshev polynomials in the range $130$-$136$ and an accuracy of eight decimal places.}
\label{DominancemL02}
\end{figure}

It is worth noting that for $m^2 \leq 3 \Lambda/4$, the QNFs of pure de Sitter spacetime are purely imaginary, whereas for $m^2> 3\Lambda/4$, the QNFs acquire a real part. 
In the background of $4D$-EGB black holes, the dS modes also acquire a real part when the scalar field mass exceeds a specific value, as shown in Fig. \ref{DominancemL002} for $\tilde{\Lambda}=0.02$ and Fig. \ref{DominancemL02} for $\tilde{\Lambda}=0.20$. As pointed out in Ref. \cite{Aragon:2020tvq}, 
for this branch of modes, the longest-lived modes are always those with the lowest angular number, and no anomalous behavior is observed.

\clearpage

\subsection{Non-perturbative de Sitter branch}
\label{nemodes}

The non-perturbative dS branch is absent for $\Lambda=0$ and is characterized by the absence of an Einsteinian limit, as the QNFs diverge when the Gauss-Bonnet coupling constant $\alpha$ approaches zero. In contrast to the perturbative dS branch, the QNFs in the non-perturbative branch exhibit more significant variations with changes in $\tilde{\alpha}$. In Table \ref{NE3}, we present the lowest QNFs for a massive scalar field. We observe that for $\ell =0, 1$, the decay rate of non-perturbative dS modes increases as the rescaled mass $\tilde{m}$ increases, except for the case $\tilde{\Lambda}=0.200$, $\ell=0$ and $\tilde{m}=0.10$.

\begin {table}[H]
\caption {The lowest QNFs $\tilde{\omega}$ of the non-perturbative dS modes for massive scalar fields with $\ell=0, 1$, in the background of the $4D$ de Sitter EGB black hole are presented for $\tilde{\Lambda}= 0.02, 0.200$, different values of $\tilde{m}$, and $\tilde{\alpha}=0.89$ and $\tilde{\alpha}=0.74$, respectively. The QNFs were obtained via the pseudospectral Chebyshev method using a number of Chebyshev polynomials in the range 195-200
with ten decimal places of accuracy.
}
\label{NE3}\centering
\begin {tabular} { | c| c |c |c |c |c |}
\hline
\multicolumn{6}{|c|}{$\Lambda=0.020$\,, $\tilde{\alpha}=0.89$}\\
\hline
{} & $\tilde{m} =0$ & $\tilde{m}= 0.1$ &  $\tilde{m}=0.2$  & $\tilde{m}=0.3$ & $\tilde{m}=0.4$   \\\hline
$\tilde{\omega} (\ell=0)$ & -0.0175398125 i & -0.0179531752 i& -0.0190229823 i &-0.0206379975 i  & -0.02262738442 i
\\\hline
$\tilde{\omega} (\ell=1)$ &-0.0508439522 i & -0.0509357316 i  &  -0.0512099161 i & -0.0516631131 i  &  -0.0522898893 i
\\\hline
\multicolumn{6}{|c|}{$\Lambda=0.200$\,, $\tilde{\alpha}=0.74$}\\
\hline
{} & $\tilde{m} =0$ & $\tilde{m}= 0.1$ &  $\tilde{m}=0.2$  & $\tilde{m}=0.3$ & $\tilde{m}=0.4$   \\\hline
$\tilde{\omega} (\ell=0)$ & -0.0129742542 i & -0.0086290334  i & -0.0150546209 i & -0.0165184366 i & -0.0184052154 i
\\\hline
$\tilde{\omega} (\ell=1)$ & -0.0439690372 i & -0.0440505519 i & -0.0442940681 i & -0.0446965640 i & -0.0452532005 i
\\\hline
\end {tabular}
\end {table}

In Table \ref{NE1} we present the values of the lowest QNFs for a massless scalar field of the non-perturbative dS branch,  within the near extremal limit where the Cauchy horizon $\tilde{r}_C$ approaches the event horizon $\tilde{r}_H$. This corresponds to small values of $\Delta= \tilde{r}_H - \tilde{r}_C$, a regime characterized by purely imaginary frequencies. We observe that the decay rate decreases as the black hole approaches extremality. It is worth mentioning that in this limit, this branch of modes for a massless scalar field is well approximated by \cite{Cardoso:2017soq}
\begin{equation}
\label{new}
\omega_{NE} = - i (\ell + n + 1) \kappa_C = - i (\ell +n+1) \kappa_h  \,,
\end{equation}
where $\kappa_C = \frac{1}{2} |f'(r_C) |$ and $\kappa_H = \frac{1}{2} f'(r_H)$ denote the surface gravities at the Cauchy and event horizons, respectively.

\begin {table}[ht]
\caption {The lowest QNF $\tilde{\omega}$ of the non-perturbative dS branch in the near-extremal limit for massless scalar fields with $\ell=0$ in the background of $4D$ de Sitter EGB black holes. The results are shown for $\tilde{\Lambda}= 0.02$ and $0.200$, considering various values of $\tilde{\alpha}$. The QNFs were computed using the pseudospectral Chebyshev method, using a number of Chebyshev polynomials in the range of 195-200
with ten decimal places of accuracy, except for $\tilde{\alpha}=0.97$, where the accuracy is limited to eight decimal places. The analytical frequencies $\tilde{\omega}_{NE}$ are derived from Eq. (\ref{new}) evaluated at the Cauchy horizon radius $\tilde{r}_C$ and are scaled by the event horizon radius. In some cases, only four decimal places of accuracy were achievable for the QNFs.}
\label{NE1}\centering
\begin {tabular} { |c|c|c|c|c|}
\hline
\multicolumn{5}{|c|}{$\tilde{\Lambda}=0.020$}\\
\hline
$\tilde{\alpha}$ & $\tilde{r}_C$   & $\Delta= \tilde{r}_H - \tilde{r}_C$  & $\tilde{\omega}$ & $\tilde{\omega}_{NE}$  \\\hline
 0.89  & 0.906  & 0.094 & -0.0175398125 i  & -0.0175160673 i
\\\hline
 0.91 & 0.927 & 0.073 & -0.0132211649 i  & -0.0131552717 i
\\\hline
 0.93 & 0.948  & 0.052  & -0.0091524329 i &  - 0.0090531515 i
\\\hline
0.95 &  0.969 & 0.031 & -0.0053209310 i & -0.0051918527 i
\\\hline
 0.97 & 0.990 & 0.010 & -0.00171751 i & -0.001555048 i
\\\hline
\end {tabular}
\begin {tabular} { | c |c |c |c |c |c |}
\hline
\multicolumn{5}{|c|}{$\tilde{\Lambda}=0.200$}\\
\hline
$\tilde{\alpha}$ & $\tilde{r}_C$   & $\Delta= \tilde{r}_H - \tilde{r}_C$  & $\tilde{\omega}$ & $\tilde{\omega}_{NE}$  \\\hline
 0.74 & 0.904  & 0.096 &  -0.0129742542 i &  -0.0121677371 i
\\\hline
0.76 & 0.935 & 0.065 &  -0.0083157895 i & -0.0070841614 i
\\\hline
 0.77  & 0.951 & 0.049 & -0.0046587971 i &  -0.00627775 i 
\\\hline
0.78  & 0.967 & 0.033 & -0.0024050031 i &  -0.00410083 i
\\\hline
0.79  & 0.983  & 0.017 & -0.0003140900 i &  -0.00208805 i 
\\\hline
\end {tabular}
\end {table}

It is worth mentioning that, in this analysis, we have considered the non-perturbative dS modes in the near-extremal limit as the Cauchy and event horizons coalesce, following the approach outlined in Ref. \cite{Cardoso:2017soq}. However, for the Schwarzschild-dS black hole, such coalescence does not occur because the Cauchy horizon is always negative and lacks physical significance. Consequently, a comparison with the Schwarzschild-dS black hole, as we have done for other branches of modes, is not possible. The near-extremal modes for Schwarzschild-dS black holes were analyzed in Ref. \cite{Cardoso:2003sw}, where the convergence between the event and the cosmological horizons was considered, a factor not addressed in this study.

\section{Behavior of decay rates of the different branches}
\label{Dom}

\begin{figure}[h]
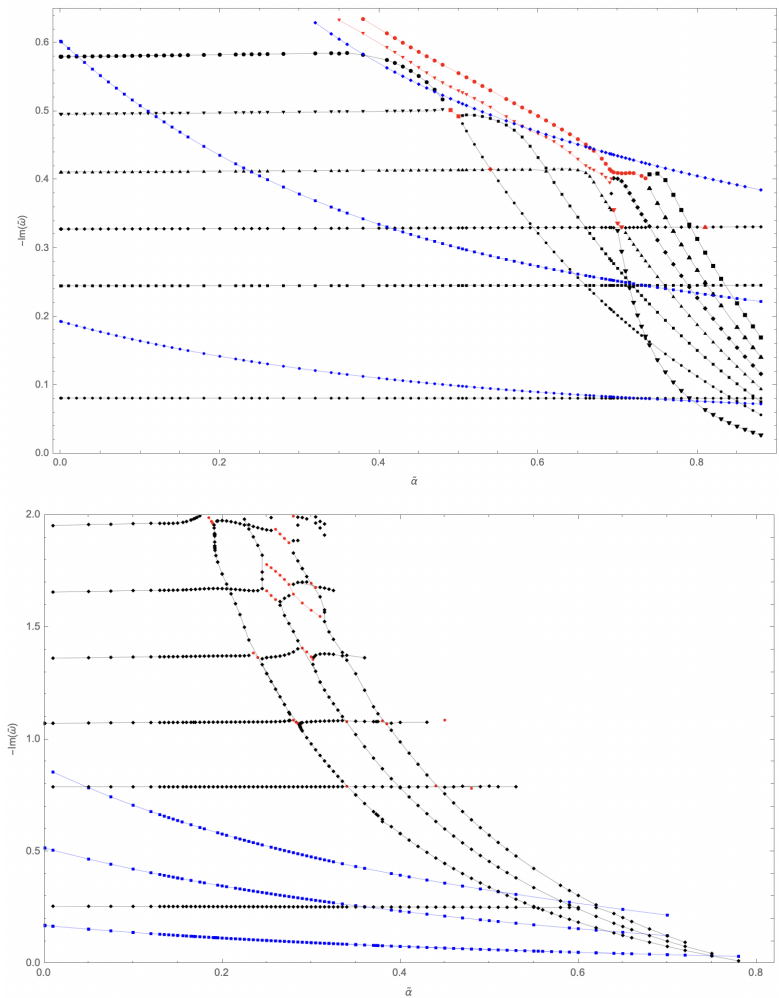

\begin{center}
\includegraphics[width=0.6\textwidth]{Dominance002l1A.pdf}
\includegraphics[width=0.595\textwidth]{Dominance02l1A.pdf}
\end{center}
\caption{The behavior of $-Im(\tilde{\omega})$ as a function of $\tilde{\alpha}$ for massless scalar field with  $\ell=1$  and $\tilde{\Lambda}=0.02$ (top panel) and $\tilde{\Lambda}=0.2$ (bottom panel).
The QNFs of the perturbative Schwarzschild branch (blue points connected by lines) correspond to complex frequencies, while the perturbative dS and non-perturbative dS branches (black points connected by lines) are associated with purely imaginary frequencies. Red points indicate complex frequencies formed by the combination of two purely imaginary frequencies of the dS branches. Conversely, a red point also can split into two purely imaginary frequencies. The QNFs were computed using the pseudospectral Chebyshev method with $155$-$150$ Chebyshev polynomials, achieving eight decimal places of accuracy. For clarity, several QNFs in each branch were omitted from the plots.}
\label{Dominance1}
\end{figure}

The imaginary part of the QNF determines the decay rate of the mode.  To analyze the decay behavior of the different branches, we plot in Fig. \ref{Dominance1} the dependence of $-Im(\tilde{\omega})$ on the coupling constant $\tilde{\alpha}$ for $\ell=1$, considering two values of the rescaled cosmological constant: $\tilde{\Lambda}=0.02$ (top panel) and $\tilde{\Lambda}=0.2$ (bottom panel). In the top panel, the fundamental modes of the dS branch initially exhibit a lower decay rate compared to the other branches at small $\tilde{\alpha}$. Around $\tilde{\alpha} \sim 0.725$, a transition occurs: the perturbative Schwarzschild branch becomes the one with the lowest decay rate (i.e. the longest-lived branch). A second transition occurs near $\tilde{\alpha} \sim 0.793$, where the non-perturbative dS branch takes over. In contrast, in the bottom panel, the perturbative Schwarzschild branch starts as the one with the lowest decay rate at small $\tilde{\alpha}$. A single transition occurs around $\tilde{\alpha} \sim 0.75$, where the non-perturbative dS branch becomes the longest-lived, with no interchange observed between the perturbative Schwarzschild and perturbative dS branches. 

\begin{figure}[h]
\begin{center}
\includegraphics[width=0.4\textwidth]{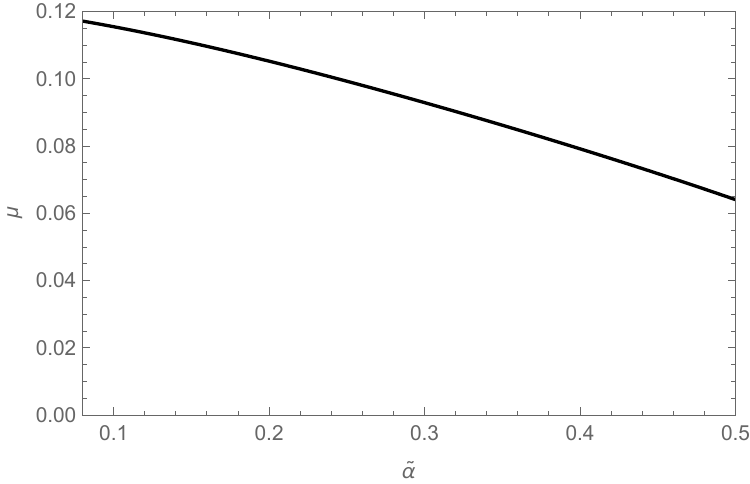}
\includegraphics[width=0.39\textwidth]{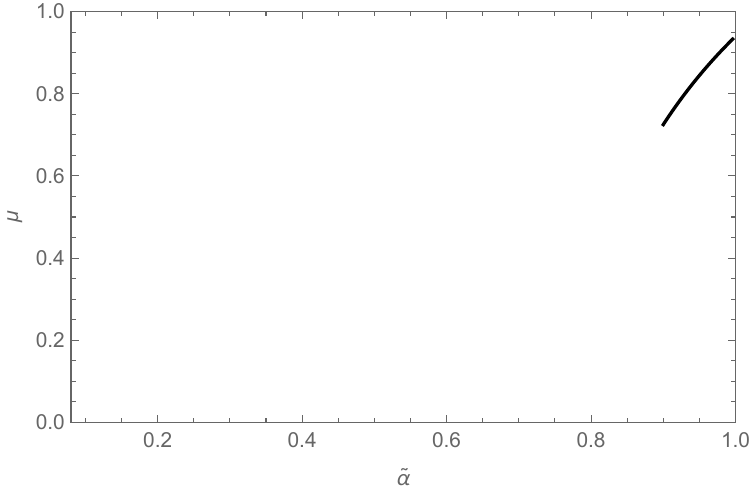}
\end{center}
\caption{The behavior of $\mu$ as a function of $\tilde{\alpha}$ for $\tilde{\Lambda}=0.02$ and $\ell=1$. Here, $\mu \approx 0.11695 $ for $\tilde{\alpha}=0.1$. The curve corresponds to $Im(\tilde{\omega}_{pure-dS})=Im(\tilde{\omega}_{PS})$ in the left panel, while in the right panel the curve corresponds to $Im(\tilde{\omega}_{PS}) = Im(\tilde{\omega}_{NE})$.}
\label{Dominance}
\end{figure}

First, it is important to note that for a fixed value of $\tilde{\alpha}$, there exists a value of $\tilde{\Lambda}$ where an interchange occurs in the lowest decay rate --from the perturbative dS branch to the perturbative Schwarzschild branch-- as shown in Table \ref{DL}. As $\tilde{\Lambda}$ increases, the non-perturbative dS modes eventually become the ones with the lowest decay rate.
\begin {table}[H]
\caption {The lowest QNFs $\tilde{\omega}$ for a massless scalar field with $\ell=1$ in the background of $4D$ de Sitter EGB black holes. The QNFs were obtained via the pseudospectral Chebyshev method with 95-100 Chebyshev polynomials, achieving ten decimal places of accuracy.}
\label{DL}\centering
\begin {tabular} { | c| c |c |c |c |c |}
\hline
\multicolumn{6}{|c|}{$\tilde{\alpha}=0.5$}\\
\hline
{} & $\tilde{\Lambda} =0.020$ & $\tilde{\Lambda}= 0.025$ & $\tilde{\Lambda}= 0.028$ &  $\tilde{\Lambda}= 0.029$ &$\tilde{\Lambda}=0.030$      
\\\hline
$\tilde{\omega} (\ell=1)$ & -0.0814645152 i  & -0.0910290241 i  & -0.0963037679 i & -0.0979974584 i   & 0.4011280031 - 0.0978825470 i 
\\\hline
\end {tabular}
\end {table}
Second, for a fixed $\tilde{\Lambda}$, the fundamental modes of the dS branch exhibit the lowest decay rate at small $\tilde{\alpha}$. However, as $\tilde{\alpha}$ increases, the perturbative Schwarzschild branch becomes the one with the longest-lived modes. Following Ref. \cite{Becar:2023jtd}, in order to obtain an approximate value of $\tilde{m}=\mu$, at which this interchange occurs, we consider the condition $Im(\tilde{\omega}_{dS})=Im(\tilde{\omega}_{PS})$ as a proxy for the transition point, where $\tilde{\omega}_{PS}$ is approximated by the WKB method at third order beyond the eikonal limit, and $\tilde{\omega}_{dS}$ is given by the analytical formula given in Eq. (\ref{deS}) for pure de Sitter spacetime, $\omega_{pure-dS}$, with an effective cosmological constant: $\Lambda_{eff}=-\frac{3}{2 \tilde{\alpha}} \left(1 - \sqrt{1 + \frac{4 \tilde{\alpha} \tilde{\Lambda}}{3}}\right)
$. This approach provides accurate approximations for the QNFs of the dS family, particularly for large $\ell$ or small $\tilde{\alpha}$. This approach is important because it helps to determine whether the branch with the longest-lived modes undergoes the anomalous behavior of the decay rate. Although an analytical expression for $\mu$ is not available for the background considered, we use the following parameters: $\tilde{\Lambda}=0.02$, $\ell=1$, and $n_{PS}=n_{dS}=0$.
In Fig. \ref{Dominance} (left panel), the curve where $Im(\omega_{pure-dS})=Im(\tilde{\omega}_{PS})$ separates regions according to the branch with the lowest decay rate, using the analytical approximation. Note that $\mu$ decreases as the GB parameter increases for the chosen parameters. Below the curve, the dS modes have the lowest decay rate, while above it, the perturbative Schwarzschild modes take over. This behavior is further confirmed in Table \ref{Dq} for $\tilde{\alpha} = 0,\, 0.1$, and $\ell=1$, where the purely imaginary QNFs belong to the perturbative dS branch, and complex QNFs to the perturbative Schwarzschild branch.

Additionally, we observe that when the perturbative Schwarzschild modes have the lowest decay rate for a massless scalar field, the same behavior occurs for a massive scalar field with $\ell=2$. For $\ell>2$, the modes with the lowest decay rates always correspond to the perturbative Schwarzschild modes for the parameters considered.  In Fig. \ref{Dominance} (right panel), the condition $Im(\tilde{\omega}_{PS}) = Im(\tilde{\omega}_{NE})$ is a proxy that marks the transition between the lowest decay rate of the perturbative Schwarzschild modes and the non-perturbative dS modes. Here, $\tilde{\omega}_{PS}$ is approximated by the WKB method at third order beyond the eikonal limit, and $\tilde{\omega}_{NE}$ is approximated by the analytical expression given in Eq. (\ref{new}). To the left of the curve, the modes with the lowest decay rate correspond to the perturbative Schwarzschild branch, whereas to the right, the non-perturbative dS modes become the longest-lived.

\begin {table}[h]
\caption {The fundamental QNFs $\tilde{\omega}$ for a massive scalar field with $\ell=1,2$ in the background of $4D$ de Sitter EGB black holes, with $\tilde{\Lambda}= 0.02$, $\tilde{\alpha}= 0, 0.1$. Here, the QNFs were obtained via the pseudospectral Chebyshev method using 95-100 Chebyshev polynomials with ten decimal places of accuracy.}
\label {Dq}\centering
\resizebox{1.0\textwidth}{!}{
\begin {tabular} { | c |c |c |c |c |}
\hline
\multicolumn{5}{ |c| }{$\tilde{\alpha}=0$}\\\hline
$\ell=1$ & $\tilde{m}=0.1$ &  $\tilde{m}=0.11$  & $\tilde{m}=0.14$ & $\tilde{m}=0.15$ \\\hline
$\tilde{\omega}$ & -0.1327316087 i  & -0.1498562721 i  & $\pm$ 0.5784401135-0.1911639971 i  & $\pm$ -0.5790851213-0.1907897675 i\\\hline
$\ell=2$ & $\tilde{m}=0$ &  $\tilde{m}=0.2$  & $\tilde{m}=0.4$ & $\tilde{m}=0.6$ \\\hline
$\tilde{\omega}$ & -0.1632904022 i & $\pm$ 0.9563702629-0.1889890368 i & $\pm$ 0.9750448286-0.1828343540 i  & $\pm$ 1.0064756591-0.1724114718 i
\\\hline
\multicolumn{5}{ |c| }{$\tilde{\alpha}=0.1$}\\\hline
$\ell=1$ & $\tilde{m}=0.1$ &  $\tilde{m}=0.11$  & $\tilde{m}=0.14$ & $\tilde{m}=0.15$ \\\hline
$\tilde{\omega}$ & -0.1326698739 i  &  -0.1498502089 i  & $\pm$  0.5400568086-0.1627721842 i & $\pm$ 0.5407261360-0.1624312502 i \\\hline
$\ell=2$ & $\tilde{m}=0$ &  $\tilde{m}=0.2$  & $\tilde{m}=0.4$ & $\tilde{m}=0.6$ \\\hline
$\tilde{\omega}$ & $\pm$ 0.8854634148-0.1630747630 i  & $\pm$ 0.8917689744-0.1612414628 i & $\pm$ 0.9108129247-0.1556603353 i & $\pm$ 0.9430015887-0.1460678570 i
\\\hline
\end {tabular}
}
\end {table}

\newpage

\section{Final Remarks}
\label{conclusion}

In this work, we considered 4D Einstein-Gauss-Bonnet black holes in dS spacetime as backgrounds and studied the propagation of massive scalar fields. We analyzed the QNMs, using both the pseudospectral Chebyshev method and the WKB method, and identified three distinct branches of modes: The first branch is the perturbative Schwarzschild branch. The QNFs of this branch are complex and smoothly reduce to the QNFs of the Schwarzschild-dS spacetime in the limit $\tilde{\alpha} \rightarrow 0$. The second branch corresponds to the perturbative dS branch, with purely imaginary QNFs for small scalar field masses and small $\tilde{\alpha}$. However, the frequencies acquire a real part as the scalar field mass increases, a behavior similar to that observed in the pure de Sitter background. These modes exhibit small deviations from the corresponding QNMs of the Schwarzschild-dS limit as $\alpha$ increases. The third branch corresponds to the non-perturbative dS branch, characterized by purely imaginary frequencies. This branch does not exist in the limit $\tilde{\alpha} \rightarrow 0$ and is thus non-perturbative in $\alpha$. In contrast to the perturbative dS branch, the QNFs in the non-perturbative branch exhibit more significant variations with changes in $\tilde{\alpha}$. The purely imaginary frequencies from both dS branches can merge, resulting in complex frequencies making them indistinguishable from each other. The propagation of scalar fields in this background leads to stable branches under the considered parameter values.

{\bf Anomalous decay rate}: For massless scalar fields, we observed that the perturbative Schwarzschild modes exhibit an anomalous decay behavior, where the decay rate decreases with increasing $\ell$, making the higher $\ell$ modes longer-lived. In other words, 
the absolute value of the imaginary part of QNFs decreases as the angular quantum number $\ell$ increases. However, when the scalar field acquires mass, this trend reverses beyond a critical mass value, with higher $\ell$ modes decaying faster. This behavior has also been observed in other dS spacetimes. The critical mass $\tilde{m}_c$ depends on both $\tilde{\Lambda}$ and $\tilde{\alpha}$. Specifically, $\tilde{m}_c$ increases with $\tilde{\Lambda}$. Additionally, based on the behavior observed in Fig. \ref{mcex1}  (left panel), the critical mass exhibits a non-monotonic dependence on $\tilde{\alpha}$: it initially decreases with increasing $\tilde{\alpha}$, reaching a minimum, and then starts to increase.


On the other hand,  the frequency of oscillation increases with $\ell$. Additionally, as the GB coupling constant increases, the frequency of oscillation also increases, while the absolute value of the imaginary part of the QNFs  decreases. Therefore, the effect of the GB term is to increase the frequency of oscillation and increase the decay time of the modes, compared to the  Schwarzschild dS spacetime. 



{\bf Branch Dominance and Transitions:} The dominance of a particular QNM branch, in terms of having the lowest decay rate, varies with the parameters. We have shown that for a massless scalar field, small values of the angular momentum ($\ell=0,1$) and small $\tilde{\Lambda}$ of the perturbative dS branch have fundamental modes with the lowest decay rate. As $\tilde{\alpha}$ or $\tilde{m}$ increases, the perturbative Schwarzschild branch becomes the one with the longest-lived fundamental modes. Furthermore, we found that the perturbative Schwarzschild modes become the longest-lived as the angular momentum $\ell$ increases. As the black hole approaches the extremal limit, the non-perturbative dS modes become the longest-lived.


\acknowledgments

We thank the referee for his/her constructive comments as well as for useful comments and suggestions. This work is partially supported by ANID Chile through FONDECYT Grant Nº 1220871  (P.A.G., and Y. V.). P. A. G. acknowledges the hospitality of the Universidad de La Serena, and  R.B acknowledges the hospitality of the Universidad Diego Portales where part of this work was undertaken.

\appendix{}

\section{A comparative analysis of the Pseudospectral Chebyshev and WKB methods}
\label{Accuracy}

In Table \ref{TableI} we show the fundamental QNFs and the first overtone, calculated using both the pseudospectral Chebyshev method and 
the sixth-order WKB method with Padé approximants. We observe a good agreement between the two methods for high values of $\ell$. Additionally, we show the relative error of the real and imaginary parts of the QNFs obtained using the WKB method, compared to the values obtained with the pseudospectral Chebyshev method. The relative error is defined by
\begin{eqnarray}
\label{E}
\epsilon_{Re(\omega)} &=& \frac{\mid Re(\omega) - Re(\omega)_{\text{WKB}} \mid}{\mid Re(\omega) \mid } \cdot 100\%\,, \\
\epsilon_{Im(\omega)} &=&\frac{\mid Im(\omega) - Im(\omega)_{\text{WKB}}\mid}{\mid Im(\omega)\mid } \cdot 100\% \,,
\end{eqnarray}
where $\omega$ denotes the result obtained using the pseudospectral Chebyshev method, and $\omega_{WKB}$ corresponds to the result obtained with the WKB method.  We observe that the error does not exceed 0.789
$\%$ in the imaginary part and 1.798
$\%$ in the real part for low values of $\ell=0,1,2,3,5$. However, for higher values of $\ell$ ($\ell = 10, 15$), the error is extremely small, not exceeding 
2.037$\cdot$10$^{-5}$ $\%$
in the imaginary part and 
1.619$\cdot$10$^{-7}$ $\%$
in the real part. Also, as observed, all frequencies have a negative imaginary part, indicating that the propagation of massive scalar fields is stable in this background.

{\bf{
\begin{table}[H]
\centering
\caption{The fundamental ($n=0$) QNFs ($\tilde{\omega}$) for several values of the angular momentum $\ell$ of the scalar field with $\tilde{m}=0$, for $4D$ de Sitter EGB black holes with $\tilde{\alpha}=0.1$, are calculated using the pseudospectral Chebyshev method and the sixth-order WKB method with Padé approximants. The QNFs obtained via the pseudospectral Chebyshev method have been calculated using a number of Chebyshev polynomials in the range $95$-$100$, with night decimal places of accuracy.}
\begin{tabular}{|c|c|c|c|c|}  \hline
\multicolumn{5}{|c|}{$\tilde{\Lambda}=0.02$}  \\  \hline
$\ell$ &  Pseudospectral Chebyshev method & WKB & $\epsilon_{Re(\tilde{\omega})}(\%)$ & $\epsilon_{Im(\tilde{\omega})}(\%)$ \\
\hline

0  &$\pm$ 0.203720441 - 0.178696218 i
 &$\pm$ 0.207383461 - 0.177286602 i & 1.798 & 0.789 \\

  1  & $\pm$ 0.535541699 - 0.165061192 i
 & $\pm$ 0.534856047 - 0.166276859 i & 0.128 & 0.736  \\
 
  2  & $\pm$ 0.885463415 - 0.163074763 i
 & $\pm$ 0.885401794 - 0.163189149 i & 6.959$\cdot 10^{-3}$ & 7.014$\cdot 10^{-2}$ \\

  3   & $\pm$ 1.237209805 - 0.162510325 i
 & $\pm$ 1.237201508 - 0.162528960 i & 6.706$\cdot 10^{-4}$ & 1.147$\cdot 10^{-2}$ \\

    5  & $\pm$ 1.941939926 - 0.162162260 i & $\pm$ 1.941939469 - 0.162163702 i & 2.353$\cdot 10^{-5}$ & 8.892$\cdot 10^{-4}$\\

   10  & $\pm$ 3.705301103 - 0.161992506 i & $\pm$ 3.705301097 - 0.161992539 i & 1.619$\cdot 10^{-7}$ & 2.037$\cdot 10^{-5}$  \\

  15  & $\pm$ 5.469124977 - 0.161958072 i & $\pm$ 5.469124977 - 0.161958075 i & 0 & 1.852$\cdot 10^{-6}$ \\
\hline
\multicolumn{5}{|c|}{$\tilde{\Lambda}=0.20$} \\ \hline

0  &$\pm$ 0.148927985 - 0.179580248 i
 & $\pm$ 0.143785803 - 0.178800090 i & 3.453 & 0.434 \\

  1  & $\pm$ 0.423826152 - 0.140523221 i
  & $\pm$ 0.423681359 - 0.140965265 i & 3.416$\cdot 10^{-2}$ & 0.315  \\

  2  & $\pm$ 0.721080268 - 0.135058398 i
 & $\pm$ 0.721062932 - 0.135100082 i &  2.404$\cdot 10^{-3}$ & 3.086$\cdot 10^{-2}$ \\

  3   & $\pm$ 1.015623674 - 0.133651268 i
 & $\pm$ 1.015621312 - 0.133657855 i  &  2.326$\cdot 10^{-4}$ & 4.928$\cdot 10^{-3}$  \\

    5  & $\pm$ 1.602049302 - 0.132805053 i  & $\pm$ 1.602049175 - 0.132805547 i & 7.927$\cdot 10^{-6}$ & 3.720$\cdot 10^{-4}$  \\

   10  & $\pm$ 3.064239155 - 0.132395269 i & $\pm$ 3.064239154 - 0.132395280 i   & 3.263$\cdot 10^{-8}$ & 8.308$\cdot 10^{-6}$  \\

  15  & $\pm$ 4.525153957 - 0.132312134 i & $\pm$ 4.525153957 - 0.132312135 i & 0 &  7.558$\cdot 10^{-7}$\\
\hline

\end{tabular}
\label{TableI}
\end{table}

}}

\section{Critical mass}
\label{wkba}

In this appendix we obtain an analytical expression for the critical scalar field mass using the WKB method up to third order beyond the eikonal limit.
\newline

Defining $L^2= \ell (\ell+1)$, we find that for large values of $L$, the maximum of the effective potential is approximately at
\begin{equation}
\notag   \tilde{r}_{max}= \tilde{r}_{0}+\frac{\tilde{r}_{1}}{L^{2}}\,,
\end{equation}
where
\begin{equation}
\notag \tilde{r}_{0}=\frac{C^{2/3} + 27 (1 + \tilde{\alpha})^2 + 18 (-1 + \tilde{\alpha} + 2 \tilde{\alpha}^2 (2 + \tilde{\alpha})) \tilde{\Lambda} - 3 (-1 + 8 \tilde{\alpha} (1 + \tilde{\alpha})) \tilde{\Lambda}^2 + 4 \tilde{\alpha} \tilde{\Lambda}^3}{2 C^{1/3} (3 + 4 \tilde{\alpha} \tilde{\Lambda})}\,,
\end{equation}

\begin{eqnarray}
\notag C&=&-16 \tilde{\alpha} F (3 + 4 \tilde{\alpha} \tilde{\Lambda})^2 \\
\notag &&+ \sqrt{-F^2 (3 + 4 \tilde{\alpha} \tilde{\Lambda})^3 \left(81 \tilde{\alpha}^4 + 4 \tilde{\alpha}^3 (81 - 283 \tilde{\Lambda}) - 12 \tilde{\alpha} (-3 + \tilde{\Lambda})^3 + (-3 + \tilde{\Lambda})^4 + 6 \tilde{\alpha}^2 \left(-47 + 9 (-6 + \tilde{\Lambda}) \tilde{\Lambda} \right) \right)}\,.
\end{eqnarray}

Also, $\tilde{r}_{1}$ can be expressed in terms of $\tilde{r}_{0}$ as
\begin{eqnarray}
\notag  \tilde{r}_{1}=&& \frac{36 \tilde{r}_0 \tilde{\alpha}^3F^2 - 2 \tilde{r}_0^3 \tilde{\alpha}^2 (15 + 2 \sqrt{3} \sqrt{B} + 12 \tilde{m}^2 \tilde{\alpha}) F^2 - 18 \tilde{r}_0^4 \tilde{\alpha}^2 F (3 + 4 \tilde{\alpha} \tilde{\Lambda})}{12 \sqrt{3} \tilde{\alpha}^2 H} \nonumber \\
\notag && - \frac{2 \tilde{r}_0^9 (3 + 4 \tilde{\alpha} \tilde{\Lambda}) (3 (2 + \tilde{m}^2 \tilde{\alpha}) (3 + 4 \tilde{\alpha} \tilde{\Lambda}) - \sqrt{3} \sqrt{B} (6 + 3 \tilde{m}^2 \tilde{\alpha} + 4 \tilde{\alpha} \tilde{\Lambda}))}{12 \sqrt{3} \tilde{\alpha}^2 H} \nonumber \\
\notag && + \frac{\tilde{r}_0^6 \tilde{\alpha} F (-3 (23 + 10 \tilde{m}^2 \tilde{\alpha}) (3 + 4 \tilde{\alpha} \tilde{\Lambda}) + \sqrt{3} \sqrt{B} (45 + 24 \tilde{m}^2 \tilde{\alpha} + 28 \tilde{\alpha} \tilde{\Lambda}))}{12 \sqrt{3} \tilde{\alpha}^2 H}\,,
\end{eqnarray}

where
\begin{equation}
\notag H=-12 \sqrt{B} \tilde{r}_0 \tilde{\alpha} F + 5 \sqrt{3} \tilde{\alpha} F^2 + 2 \sqrt{3} F \tilde{r}_0^3 (3 + 4 \tilde{\alpha} \tilde{\Lambda}) - 3 \sqrt{B} \tilde{r}_0^4 (3 + 4 \tilde{\alpha} \tilde{\Lambda})\,,
\end{equation}
\begin{equation}
\notag F=3+3\tilde{\alpha}-\tilde{\Lambda}\,. 
\end{equation}

\begin{equation}
\notag V(r^{*}_{max})\approx V_{0}L^{2}+V_{1}
\end{equation}
where
\begin{equation}
\notag V_{0}= \frac{L^2 ((3 - \sqrt{3} \sqrt{B}) \tilde{r}_0^2 + 6 \tilde{\alpha})}{6 \tilde{r}_0^2 \tilde{\alpha}}\,, 
\end{equation}
\begin{eqnarray}
\notag V_{1}=&& \frac{G \tilde{r}_0^2 \left(9 \tilde{\alpha}  \sqrt{B} F+\sqrt{3} \left(3-\sqrt{3} \sqrt{B}\right) \left(4 \tilde{\alpha}  \left(3 \tilde{\alpha} +\tilde{\Lambda}  \left(\tilde{r}_0^3-1\right)+3\right)+3 \tilde{r}_0^3\right)\right)}{108 \tilde{\alpha} ^3 H \tilde{r}_0^3 \left(4 \tilde{\alpha}  \left(3 \tilde{\alpha} +\tilde{\Lambda}  \left(\tilde{r}_0^3-1\right)+3\right)+3 \tilde{r}_0^3\right)}\\
\notag && -\frac{\tilde{\alpha}  \left(6 \tilde{\alpha} +\left(3-\sqrt{3} \sqrt{B}\right) \tilde{r}_0^2\right) \left(\tilde{r}_0^3 \left(4 \sqrt{3} \tilde{\alpha}  \tilde{\Lambda} -3 \tilde{\alpha}  \sqrt{B} \tilde{m}^2-3 \sqrt{B}+3 \sqrt{3}\right)+\sqrt{3} \tilde{\alpha}  F\right)}{18 \tilde{\alpha} ^3 \sqrt{B} \tilde{r}_0^3}\\
\notag && -\frac{\sqrt{3} G \left(6 \tilde{\alpha} +\left(3-\sqrt{3} \sqrt{B}\right) \tilde{r}_0^2\right)}{108 \tilde{\alpha} ^3 H \tilde{r}_0^3}+\frac{L^2 \left(6 \tilde{\alpha} +\left(3-\sqrt{3} \sqrt{B}\right) \tilde{r}_0^2\right)}{6 \tilde{\alpha}  \tilde{r}_0^2}\,,
\end{eqnarray}
\begin{eqnarray}
\notag G=&& 36 \tilde{r}_0 \tilde{\alpha}^3 F^2 - 2 \tilde{r}_0^3 \tilde{\alpha}^2 (15 + 2 \sqrt{3} \sqrt{B} + 12 \tilde{m}^2 \tilde{\alpha}) F^2 \nonumber  - 18 F \tilde{r}_0^4 \tilde{\alpha}^2 (3 + 4 \tilde{\alpha} \tilde{\Lambda}) + F \tilde{r}_0^6 \tilde{\alpha} (45 \sqrt{3} \sqrt{B} + 24 \sqrt{3} \sqrt{B} \tilde{m}^2 \tilde{\alpha} \nonumber \\
\notag && + 28 \sqrt{3} \sqrt{B} \tilde{\alpha} \tilde{\Lambda} - 3 (23 + 10 \tilde{m}^2 \tilde{\alpha}) (3 + 4 \tilde{\alpha} \tilde{\Lambda})) \nonumber  - 2 \tilde{r}_0^9 (3 + 4 \tilde{\alpha} \tilde{\Lambda}) (18 - 6 \sqrt{3} \sqrt{B} + 24 \tilde{\alpha} \tilde{\Lambda} - 4 \sqrt{3} \sqrt{B} \tilde{\alpha} \tilde{\Lambda} \nonumber \\
\notag && + 3 \tilde{m}^2 \tilde{\alpha} (3 - \sqrt{3} \sqrt{B} + 4 \tilde{\alpha} \tilde{\Lambda}))\,,
\end{eqnarray}
while the higher order derivatives $V^{(i)}(r^{*}_{max})$ for $i=2,...,6$, can be expressed in the following abbreviated  manner
\begin{align}
 \notag   V^{(2)}(r^{*}_{max})&\approx V_{0}^{(2)}L^{2}+V_{1}^{(2)}\\
 \notag   V^{(3)}(r^{*}_{max})&\approx V_{0}^{(3)}L^{2}\\
\notag    V^{(4)}(r^{*}_{max})&\approx V_{0}^{(4)}L^{2}\\
 \notag   V^{(5)}(r^{*}_{max})&\approx V_{0}^{(5)}L^{2}\\
 \notag   V^{(6)}(r^{*}_{max})&\approx V_{0}^{(6)}L^{2}~.
\end{align}
Nevertheless, the coefficients of each of the equations are too lengthy to be presented here.

On the other hand, our interest is to evaluate the QNFs for large values of $L$, so we expand the frequencies as a power series in $L$. It is important to keep in mind that in the eikonal limit, the leading term is linear in $L$, and for $\tilde{\alpha}=0$, we should recover the frequencies of the Schwarzschild-de Sitter black hole. Next, we consider the following expression in powers of $L$
\begin{equation}
\label{omegawkb}
\tilde{\omega}=\tilde{\omega}_{1m}L+\tilde{\omega}_{0}+\tilde{\omega_{1}}L^{-1}+\tilde{\omega}_{2}L^{-2} + \mathcal{O}(L^{-3})\,,
\end{equation}
where
\begin{equation}
\notag 
 \tilde{\omega}_{1m}=\sqrt{\frac{1}{\tilde{r}_0^2} + \frac{1}{6\tilde{\alpha}}\left(3 - \sqrt{3} \sqrt{3 + \frac{4\tilde{\alpha}F}{\tilde{r}_0^3} + 4\tilde{\alpha}\tilde{\Lambda}}\right)}\,,
\end{equation}

\begin{align*}
\tilde{\omega}_{0}=& -\frac{1}{2 \cdot 3^{3/4} \tilde{\omega}_{1m}} i \left( \frac{1}{2} + n \right)   \sqrt{ \frac{ \left(-6 \tilde{\alpha} + \tilde{r}_0^2 \left(-3 + \sqrt{3} B\right) \right) }{ \tilde{r}_0^5 \tilde{\alpha}^2 B \left( 4 \tilde{\alpha} F + \tilde{r}_0^3 (3 + 4 \tilde{\alpha} \tilde{\Lambda}) \right) }}  \Bigg( -70 \tilde{\alpha}^2 F^2 + 24 \sqrt{3} \tilde{r}_0 \tilde{\alpha}^2 F B \\
&  + 6 \sqrt{3} \tilde{r}_0^4 \tilde{\alpha} (3 + 4 \tilde{\alpha} \tilde{\Lambda}) B  - 2 \tilde{r}_0^3 \tilde{\alpha} F (39 + 52 \tilde{\alpha} \tilde{\Lambda} - 2 \sqrt{3} B)+ \tilde{r}_0^6 (3 + 4 \tilde{\alpha} \tilde{\Lambda}) (-3 - 4 \tilde{\alpha} \tilde{\Lambda} + \sqrt{3} B) \\
&  + \tilde{r}_0^2 \tilde{\alpha} F^2 (-3 + 4 \sqrt{3} B) + \tilde{r}_0^5 (-3 + \sqrt{3} B) (9 - 3 \tilde{\Lambda} + 12 \tilde{\alpha}^2 \tilde{\Lambda} + \tilde{\alpha} (9 + 12 \tilde{\Lambda} - 4 \tilde{\Lambda}^2)) \Bigg)^{\frac{1}{2}}\,, 
\end{align*}
where $B=\sqrt{3+\frac{4\tilde{\alpha}F}{\tilde{r}_{0}^{3}}+4\tilde{\alpha}\tilde{\Lambda}}$. Therefore, for small values of the parameter $\tilde{\alpha}$, the critical mass $\tilde{m}_{c}$ can be approximated by expanding it in terms of $\tilde{\alpha}$ up to the fourth order as follows
\begin{equation}
\label{mcritical}
\tilde{m}_{c} \approx m_0 + \tilde{\alpha} m_1 + \tilde{\alpha}^2 m_2+ \tilde{\alpha}^3 m_3 + \tilde{\alpha}^4 m_4 + \dots   
\end{equation}
\small{ where
\begin{eqnarray}
\notag m_0 &=& A / \Big(18 \sqrt{10} \Big) \,, \\
\notag m_1 &=& \Big( -10025 \tilde{\Lambda }^6+120300 \tilde{\Lambda }^5-541350 \tilde{\Lambda }^4+1116700 \tilde{\Lambda }^3-1016025 \tilde{\Lambda }^2+355320 \tilde{\Lambda }-526904 \Big) / \Big(  540 \sqrt{10} A \left(\tilde{\Lambda }-3\right)^4 \Big) \,, \\
\notag  m_2 &=& \Big(  703504375 \tilde{\Lambda }^{12}-16884105000 \tilde{\Lambda }^{11}+177283102500 \tilde{\Lambda }^{10}-1061873263000 \tilde{\Lambda }^9+3956013470250 \tilde{\Lambda }^8  \\ 
\notag  && -9301334949000 \tilde{\Lambda }^7+12975135918900 \tilde{\Lambda }^6-7572452779800 \tilde{\Lambda }^5-6675204148425 \tilde{\Lambda }^4+20728852642160 \tilde{\Lambda }^3- \\
\notag  && 30943720058160 \tilde{\Lambda }^2+25510630437360 \tilde{\Lambda }+497662960768  \Big) / \Big( 32400 \sqrt{10} A^3 \left(\tilde{\Lambda }-3\right)^8 \Big) \,, \\
\notag m_3 &=& -\Big(6649623853125 \tilde{\Lambda }^{18}-239386458712500 \tilde{\Lambda }^{17}+3949876568756250 \tilde{\Lambda }^{16}-39485826031092500 \tilde{\Lambda }^{15}+  \\
\notag && 266228478696946875 \tilde{\Lambda }^{14}-1274630383212570000 \tilde{\Lambda }^{13}+4439587796068616500 \tilde{\Lambda }^{12}-11319629484435351000 \tilde{\Lambda }^{11}+  \\
\notag && 20928660269052328875 \tilde{\Lambda }^{10}-27398792379281700500 \tilde{\Lambda }^9+25352052160994930250 \tilde{\Lambda }^8-23294258197195774500 \tilde{\Lambda }^7+  \\
\notag &&  46864287612917366645 \tilde{\Lambda }^6-111227858990108638440 \tilde{\Lambda }^5+174913304709136163400 \tilde{\Lambda }^4-154774535783282651056 \tilde{\Lambda }^3   \\
\notag && +57018588374842538496 \tilde{\Lambda }^2+850180523745580416 \tilde{\Lambda }+1755977419983872   \Big) / \Big( 194400 \sqrt{10} A^5 \left(\tilde{\Lambda }-3\right)^{12} \Big) \,, \\
\notag m_4 &=& \Big( 1444353714430859375 \tilde{\Lambda }^{24}-69328978292681250000 \tilde{\Lambda }^{23}+1559902011585328125000 \tilde{\Lambda }^{22}-  \\
\notag && 21835786604628925750000 \tilde{\Lambda }^{21}+212807279163639233062500 \tilde{\Lambda }^{20}-1530756581838889758750000 \tilde{\Lambda }^{19}+  \\
\notag && 8404429000471627934975000 \tilde{\Lambda }^{18}-35914278783763054445850000 \tilde{\Lambda }^{17}+120663429899892878596556250 \tilde{\Lambda }^{16}- \\
\notag && 319630980665525686572130000 \tilde{\Lambda }^{15}+665074600544465310887175000 \tilde{\Lambda }^{14}-1076093783320218852363030000 \tilde{\Lambda }^{13} \\
\notag && +1330227896397921007264902500 \tilde{\Lambda }^{12}-1204920179101751592604890000 \tilde{\Lambda }^{11}+599873276804525381275497000 \tilde{\Lambda }^{10} \\
\notag && +738631808844496177924581200 \tilde{\Lambda }^9-3479157145369928413617102225 \tilde{\Lambda }^8+7684699389084324479813839200 \tilde{\Lambda }^7-   \\
\notag &&  11179795882946229900312531040 \tilde{\Lambda }^6+10821702301832821822363541280 \tilde{\Lambda }^5-6567577183404421754583548160 \tilde{\Lambda }^4 \\
\notag && +2046482024312484129357812480 \tilde{\Lambda }^3-93099370670254254011347200 \tilde{\Lambda }^2-2084980705184254439854080 \tilde{\Lambda }  \\
\notag && -4138979521391922085888 \Big) / \Big( 23328000 \sqrt{10} A^7 \left(\tilde{\Lambda }-3\right)^{16} \Big) 
 \,,
\end{eqnarray}

}

\normalsize{

where $A =\sqrt{ 548+2005(-3+\tilde{\Lambda})^{2}\tilde{\Lambda}} /\left|\tilde{\Lambda }-3\right|$. Note that by setting the parameter $\tilde{\alpha}=0$ we recover the critical mass of the Schwarzschild-de Sitter black hole presented in \cite{Aragon:2020tvq}.

}

\end{document}